\documentclass[useAMS,usedcolumn,usenatbib,usegraphicx]{mn2e}

\def\oii{\ifmmode \rmn{[O}\,\textsc{ii}\rmn{]} \else [O\,\textsc{ii}]\fi}
\def\oiii{[O\,\textsc{iii}]}
\def\ha{\ifmmode \rmn{H}{\alpha}\else H$\alpha$\fi}
\def\hb{\ifmmode \rmn{H}{\beta}\else H$\beta$\fi}
\def\m15{\ifmmode \Delta m_{15}(B) \else $\Delta m_{15}(B)$\fi}
\def\ebmv{\ifmmode E(B-V) \else $E(B-V)$\fi}
\def\eumb{\ifmmode E(U-B) \else $E(U-B)$\fi}
\def\evmr{\ifmmode E(V-R) \else $E(V-R)$\fi}
\def\bmvmax{\ifmmode B_{\rmn{max}}-V_{\rmn{max}} \else $B_{\rmn{max}}-V_{\rmn{max}}$\fi}
\newcommand{\hii}{\hbox{H\,\textsc{ii}}}
\newcommand{\hst}{\textit{HST}}
\newcommand{\iraf}{\textsc{iraf}}
\newcommand{\splot}{\textsc{splot}}
\newcommand{\makee}{\textsc{makee}}
\newcommand{\sex}{\textsc{sextractor}}
\newcommand\om{\ifmmode \Omega_{\rmn{M}}\else $\Omega_{\rmn{M}}$\fi}
\newcommand\ok{\ifmmode \Omega_{\rmn{k}}\else $\Omega_{\rmn{k}}$\fi}
\newcommand\ol{\ifmmode \Omega_{\Lambda}\else $\Omega_{\Lambda}$\fi}
\def\aap{A\&A}                % Astronomy and Astrophysics
\def\pasp{PASP}               % Publications of the ASP
\def\aj{AJ}                   % Astronomical Journal
\def\apj{ApJ}                 % Astrophysical Journal
\def\apjl{ApJ}                % Astrophysical Journal, Letters
\def\apjs{ApJS}               % Astrophysical Journal, Supplement
\def\mnras{MNRAS}            % Monthly Notices of the RAS
\def\procspie{Proc.~SPIE}   % Proceedings of the SPIE
\def\aaps{A\&AS}              % Astronomy and Astrophysics, Supplement
\def\nat{Nature}              % Nature

\title{The Hubble Diagram of Type Ia Supernovae as a Function of Host Galaxy Morphology}
\author[M. Sullivan et al.]
{M.~Sullivan,$^{1,2}$\thanks{E-mail: mark.sullivan@durham.ac.uk}
R.~S.~Ellis,$^2$
G.~Aldering,$^3$
R.~Amanullah,$^4$
P.~Astier,$^5$
G.~Blanc,$^{3,5}$\newauthor
M.~S.~Burns,$^6$
A.~Conley,$^{3,7}$
S.~E.~Deustua,$^{3,8}$
M.~Doi,$^9$
S.~Fabbro,$^{10}$
G.~Folatelli,$^4$\newauthor
A.~S.~Fruchter,$^{11}$
G.~Garavini,$^4$
R.~Gibbons,$^3$
G.~Goldhaber,$^{3,7}$
A.~Goobar,$^4$\newauthor
D.~E.~Groom,$^3$
D.~Hardin,$^5$
I.~Hook,$^{12}$
D.~A.~Howell,$^3$
M.~Irwin,$^{13}$
A.~G.~Kim,$^3$\newauthor
R.~A.~Knop,$^{14}$
C.~Lidman,$^{15}$
R.~McMahon,$^{13}$
J.~Mendez,$^{16,17}$
S.~Nobili,$^4$
P.~E.~Nugent,$^3$\newauthor
R.~Pain,$^5$
N.~Panagia,$^{11}$
C.~R.~Pennypacker,$^3$
S.~Perlmutter,$^3$
R.~Quimby,$^3$\newauthor
J.~Raux,$^5$
N.~Regnault,$^3$
P.~Ruiz-Lapuente,$^{16}$
B.~Schaefer,$^{18}$
K.~Schahmaneche,$^5$\newauthor
A.~L.~Spadafora,$^3$
N.~A.~Walton,$^{17,13}$
L.~Wang,$^3$
W.~M.~Wood-Vasey,$^{3,7}$
N. Yasuda$^{19}$\newauthor
(The Supernova Cosmology Project)\\
$^1$ Department of Physics, University of Durham, South Road, Durham, DH1 3LE, UK\\
$^2$ California Institute of Technology, E. California Blvd, Pasadena, CA 91125, USA\\
$^3$ E. O. Lawrence Berkeley National Laboratory, 1 Cyclotron Rd., Berkeley, CA 94720, USA\\
$^4$ Department of Physics, Stockholm University, SCFAB, S-106 91 Stockholm, Sweden\\
$^5$ LPNHE, CNRS-IN2P3, University of Paris VI \& VII, Paris, France\\
$^6$ Colorado College, 14 East Cache La Poudre St., Colorado Springs, CO 80903\\
$^7$ Department of Physics, University of California Berkeley, Berkeley, CA 94720-7300, USA\\
$^8$ American Astronomical Society, 2000 Florida Ave, NW, Suite 400, Washington, DC 20009 USA\\
$^9$ Department of Astronomy and Research Center for the Early Universe, School of Science, University of Tokyo, Tokyo 113-0033, Japan\\
$^{10}$ IST, Lisbon, Portugal\\ 
$^{11}$ Space Telescope Science Institute, 3700 San Martin Drive, Baltimore, MD 21218, USA\\
$^{12}$ Department of Physics, University of Oxford, Nuclear \& Astrophysics Laboratory, Keble Road, Oxford, OX1 3RH, UK\\
$^{13}$ Institute of Astronomy, Madingley Road, Cambridge, CB3 0HA, UK \\
$^{14}$ Department of Physics and Astronomy, Vanderbilt University, Nashville, TN 37240, USA\\
$^{15}$ European Southern Observatory, Alonso de Cordova 3107, Vitacura, Casilla 19001, Santiago 19, Chile\\ 
$^{16}$ Department of Astronomy, University of Barcelona, Barcelona, Spain\\
$^{17}$ Isaac Newton Group of Telescopes, Apartado de Correos 321, E-38700 Santa Cruz de La Palma, Islas Canarias, Spain\\
$^{18}$ University of Texas, Department of Astronomy, C-1400, Austin, TX 78712, USA\\
$^{19}$ National Astronomical Observatory, Mitaka, Tokyo 181-8588, Japan\\
}
\date {Accepted ---. Received ---; in original form ---.}

\begin{document}
\label{firstpage}
\maketitle

\clearpage

\begin{abstract}
  We present new results on the Hubble diagram of distant type Ia
  supernovae (SNe\,Ia) segregated according to the type of host
  galaxy.  This makes it possible to check earlier evidence for a
  cosmological constant by explicitly comparing SNe residing in
  galaxies likely to contain negligible dust with the larger sample.
  The cosmological parameters derived from these SNe\,Ia hosted by
  presumed dust-free early-type galaxies supports earlier claims for a
  cosmological constant, which we demonstrate at $\simeq5\sigma$
  significance, and the internal extinction implied is small even for
  late-type systems ($A_B<0.2$). Thus, our data demonstrate that host
  galaxy extinction is unlikely to systematically dim distant SNe Ia
  in a manner that would produce a spurious cosmological constant. We
  classify the host galaxies of 39 distant SNe discovered by the
  Supernova Cosmology Project (SCP) using the combination of
  \textit{Hubble Space Telescope} STIS imaging, Keck-II ESI
  spectroscopy and ground-based broad-band photometry. The distant
  data are analysed in comparison with a low-redshift sample of 25
  SNe\,Ia re-calibrated according to the precepts of the SCP. The
  scatter observed in the SNe\,Ia Hubble diagrams correlates closely
  with host galaxy morphology. We find, as expected, this scatter is
  smallest for SNe\,Ia occurring in early-type hosts and largest for
  those occurring in late-type galaxies. Moreover, SNe residing in
  early-type hosts appear only $\simeq0.14\pm0.09$\,mag brighter in
  their light-curve-width-corrected luminosity than those in late-type
  hosts, implying only a modest amount of dust extinction even in the
  late-type systems. As in previous studies, these results are broadly
  independent of whether corrections based upon SN light-curve shapes
  are performed.  We also use our high-redshift dataset to search for
  morphological dependencies in the SNe light-curves, as are sometimes
  seen in low-redshift samples. No significant trends are found,
  possibly because the range of light-curve widths is too limited.
\end{abstract}


\section{Introduction}
\label{sec:introduction}

Type Ia supernovae (SNe\,Ia) have emerged as important probes of the
cosmological world-model. Although their physical properties are still
not fully understood, by exploiting an empirical relationship between
the peak intrinsic luminosity and the light-curve decay time
\citep{1993ApJ...413L.105P,1995ApJ...438L..17R,1996ApJ...473...88R,1996AJ....112.2391H,1997ApJ...483..565P},
the dispersion in their photometric properties can be reduced to
$\simeq0.17\,\rmn{mag}$, making them valuable `calibrated candles'
over a wide range in redshift \citep[e.g.][]{1996AJ....112.2398H}.

Although numerous alternative probes of the cosmological parameters
are now available
\citep{2001PhRvL..86.3475J,2001Natur.410..169P,2002ApJ...564..559D,2002MNRAS.330L..29E},
in principle the SN\,Ia Hubble diagram provides the only
\textit{direct} measure of the cosmic expansion history free from any
model assumptions. The systematic detection \citep{1997perlbook..749}
and photometric calibration of high redshift SNe\,Ia has led to two
independent SN\,Ia Hubble diagrams, published by the Supernova
Cosmology Project \citep[hereafter
SCP;][]{1997ApJ...483..565P,1998Natur.391...51P,1999ApJ...517..565P}
and the High-Redshift Supernova Search Team \citep[hereafter
HZT;][]{1998ApJ...507...46S,1998AJ....116.1009R}.  Data from both
teams convincingly reject the deceleration expected from an
Einstein--de Sitter (EdS) Universe, and provide new evidence for a
cosmic acceleration in a low mass-density Universe consistent with a
non-zero vacuum energy density.  Evidence for spatial flatness from
microwave background experiments
\citep{2000Natur.404..955D,2000ApJ...545L...1B,2001PhRvL..86.3475J}
further strengthens these conclusions \citep[see
e.g.][]{1999Sci...284.1481B}.

Such important results demand excellent supporting evidence. In
particular, it is necessary to examine the homogeneity, environmental
trends and evolutionary behaviour of SNe\,Ia. It must be remembered
that it is not yet clear whether these events are produced by a single
mechanism in a unique physical environment.  Systematic differences in
the peak magnitudes between high and low-redshift SNe, arising either
from evolutionary effects in the SNe progenitors or via subtle
differences in the environments of the low and high-redshift galaxies
in which SNe are produced, could mimic the cosmic acceleration without
necessarily destroying the small dispersion seen in existing Hubble
diagrams.

Differences between low and high-redshift SNe\,Ia might arise via
differing progenitor compositions
\citep{1999ApJ...522L..43U,2000ApJ...528..590H,2001ApJ...557..279D},
different progenitor ages \citep*{ruizlapuente97}, greater amounts of
dust in high-redshift environments, either in the host galaxy
\citep*{1998ApJ...502..177H,1999ApJ...526L..65T,2002MNRAS.332..352R}
or in the intergalactic medium
\citep{1999ApJ...525..583A,2000ApJ...532...28A}, or a dependence of
the SNe properties on host galaxy environments.  Tests for analysing
SNe by the type of their host galaxy were discussed by
\citet[][hereafter P97]{1997ApJ...483..565P}, and continued by
\citet[][hereafter P99]{1999ApJ...517..565P}. This initial
classification based on host galaxy spectra revealed no changes in the
properties of SNe located in E/S0 and spiral hosts, though only 17
hosts could be classified and little morphological information was
available.

Continuing these studies, via an examination of the dependence of SN
properties on host galaxy morphology via \textit{Hubble Space
  Telescope} (\hst) imaging and Keck spectroscopy, forms the basis of
the present analysis.  Although SNe\,Ia can occur in all types of
galaxies, disc and spheroidal stellar populations will represent
different star-formation histories, metallicities and dust content.
Thus we might expect that SNe\,Ia progenitor composition and peak
magnitudes or light-curve properties could be affected accordingly.

Much of the previous work examining the environments of SNe\,Ia has
been conducted using low-redshift samples, with only limited work at
high-redshift. However, in these low-redshift studies, two interesting
trends have already been claimed, both of which are relevant for this
present paper. The first correlates the form of the SN
\textit{light-curve} with the host-galaxy properties.
\citet{1989ApJ...345L..43F} originally found that the initial
post-maximum decline rate in the light-curve in the $B$-band may have
a smaller dispersion among SNe located in elliptical galaxies. Using
SNe\,Ia discovered via the Calan-Tololo (C-T) SN survey
\citep{1996AJ....112.2408H}, \citet{1996AJ....112.2391H} also note
that the post-maximum decline rate correlates with morphology, in the
sense that SNe with more rapidly declining light-curves (i.e.
intrinsically dimmer SNe) occur in earlier-type galaxies. In an
independent study of the CfA SN sample, \citet{1999AJ....117..707R}
find a similar result.  The various low-redshift SN samples were later
combined in a single analysis by \citet{2000AJ....120.1479H}.  They
confirm the findings of the C-T and CfA studies, and also note that
intrinsically brighter SNe tend to occur in bluer stellar environments
\citep*[see also][]{1996ApJ...465...73B}, and suggest that the
intrinsically brightest SNe occur in the least luminous galaxies.
These effects were further investigated by
\citet{2001ApJ...554L.193H}, who find that in the local Universe
over-luminous SNe arise in spiral and other late-type systems, whilst
spectroscopically under-luminous SNe tend to occur in older (E/S0)
systems.

The second suggested correlation relates the photometric properties of
a SN to its \textit{physical location} in the host galaxy.
\citet*{1997ApJ...483L..29W}, using SNe from the C-T survey as well as
11 other well-studied nearby SNe, found that SNe properties apparently
vary with projected distance from the centre of the host galaxy. They
find that SNe located more than 7.5\,kpc from the centre show around
3\,--\,4 times less scatter in maximum brightness than those within
that projected radius, with E/S0 galaxies dominating the sample at
high separations. Monte-Carlo simulations exploring the effects of
dust on radial distributions of SNe\,Ia \citep*{1998ApJ...502..177H}
similarly predict that SNe at large projected radial distances should
be practically unextinguished, whilst those at small projected
distances could suffer significant amounts of extinction. The
conclusion is that SNe at higher projected distance are a more
homogeneous group and should be better distance indicators.

In a combined C-T/CfA sample, \citet{1999AJ....117..707R} find that
SNe\,Ia with faster decline rates (i.e. intrinsically fainter SNe)
occur at larger projected distances from the host galaxy centre. The
\citet{2000AJ....120.1479H} sample has also been used by
\citet*{2000ApJ...542..588I} for a similar purpose, though importantly
using de-projected galactocentric distances rather than simple
projected distances.  This refinement suggests that the earlier
\citet{1999AJ....117..707R} trend may partly be due to the mix of E/S0
and spiral galaxies, with (intrinsically fainter) SNe hosted by E/S0
galaxies being discovered at larger galactic distances when compared
to SNe located in spiral galaxies. No significant trend is seen in SN
parameters with galactocentric distance for the E/S0 subset of the SN
host population. As elliptical galaxies show radial gradients in their
stellar properties arising mostly from trends in metallicity rather
than age, it is concluded that the intrinsic diversity in SN
parameters more likely arises from stellar population differences
rather than progenitor metallicity alone \citep{2000ApJ...542..588I}.
The only relevant study for high-redshift samples is that of
\citet*{2000ApJ...530..166H}, based on SCP IAU circulars and finding
charts. These authors examine the radial distribution of the SCP SNe
from their host galaxies, and find that they are distributed similarly
to those from low-redshift CCD-based samples.

The major difficulty in interpreting these trends lies in
disentangling all of the numerous variables (morphology, radial
position, luminosity, metallicity, dust content and star-formation
history) which define the stellar populations of galaxies. Though no
complete physical picture has yet emerged as to how the underlying
stellar population might influence SN properties in the manner
proposed, one possible hypothesis is as follows
\citep{1999ApJ...522L..43U}.

SNe\,Ia most likely originate via the thermonuclear explosion of a
carbon-oxygen (CO) white dwarf (WD) in a binary system, with the
light-curve powered by the decay of $^{56}$Ni.  Hence, WDs with a
larger $^{12}$C mass fraction will generate more $^{56}$Ni and
consequently brighter SNe.  As the oldest progenitors are likely to
have the lowest mass companions, less mass can be transferred to the
primary WD, and the resultant reduced final $^{12}$C abundance leads
to a smaller $^{56}$Ni mass, and therefore fainter events.  In this
simple picture, one expects firstly to see more variation in SN
properties in disc-based systems (where the age of the stellar
population is more diverse) than in spheroidal galaxies, and secondly
to see fewer over-luminous events in (presumably) older E/S0 systems.

So far as the implications for the use of SNe\,Ia as cosmological
probes are concerned, we are, of course, interested in possible trends
in the above correlations with redshift. Our aim is to include SNe
located at high redshift which are used in existing Hubble diagrams to
constrain the cosmological parameters. Environmental effects on the
determination of the cosmological parameters can then be more easily
explored. An outline of the paper follows. In $\S$2 we discuss the low
and high-redshift SN samples used in our analysis and describe the
method utilised to align them on to the same calibration system.
$\S$3 introduces our new dataset, based primarily on \hst\ imaging and
Keck spectroscopy, used to investigate the properties of the host
galaxies of the high-redshift sample. We discuss the classification
system for the hosts in $\S$4, and present our analyses in $\S$5 and
discussion in $\S$6. We present our conclusions in $\S$7.

\section{The Supernova Samples}
\label{sec:supernova-samples}

In order for us to explore the cosmological consequences of
possible correlations between the SN properties and those of the
host galaxies, we need samples of SNe\,Ia spanning the full range
in redshift in the Hubble diagram analyses.

For the distant SNe\,Ia, we perform the analysis on the published
sample of 42 SNe discovered by the Supernova Cosmology Project (SCP;
see P99). All SNe were discovered by the SCP and details of the
various procedures and analysis techniques, as well as spectroscopic
verification, can be found in P97 and P99.

The bulk of our sample overlaps with the P99 set but this overlap is
not complete by virtue of the random manner in which \hst\ snapshot
observations are selected. For example, only 30 of the 42 have \hst\ 
imaging providing host galaxy morphologies. P99 further restrict this
sample of 42 SNe to one of 38 high-redshift objects in the primary
cosmological fit (`fit-C' in P99) by excluding those SNe considered
`likely reddened' or which are `outliers' in terms of their stretch of
the light-curve time-scale values.  Where appropriate in this current
analysis we also note the effect of excluding these 4 objects.

At lower redshifts, we use the \citet{1996AJ....112.2408H} SNe sample
and add selected SNe from the \citet[][]{1999AJ....117..707R} local
sample. Our principal criteria in selecting SNe from these local
samples remains as in P99, namely (i) each SN must have been observed
prior to 5 days after maximum light so as to reduce any extrapolations
necessary to fit peak luminosities, and (ii) that each SN has
$cz>3000\,\rmn{km\,s^{-1}}$ to control the impact of peculiar
velocities. These requirements mean that 18 of the 29 C-T SNe (see P99
for details) and 12 of the 22 \citet[][]{1999AJ....117..707R} CfA
SNe\footnote{SN1994m, SN1994s, SN1994t, SN1995e SN1995ac, SN1995ak,
  SN1995bd, SN1996ab, SN1996bv, SN1996bl, SN1996bo, SN1996c} are
included in our low-redshift sample of 30 objects.  As in P99, the two
large stretch/residual outliers in the C-T sample (1992bo and 1992br,
both located in E/S0 systems) are excluded in the analysis, as well as
two from the CfA sample (1994t and 1995e). One further SNe whose
light-curve fit results in a unreasonably high stretch value is also
excluded (1996ab from the CfA sample). This leaves 25 low-redshift SNe
in our primary analysis sample (16 from the C-T survey, and 9 from
CfA).

It is important to ensure that the corrected peak magnitudes of the
SNe drawn from the different samples are derived in a consistent
manner. Raw observed SN peak magnitudes have an r.m.s.  dispersion of
$\la0.30\,\rmn{mag}$ \citep[e.g.][]{1996AJ....112.2391H}, but this
dispersion can be significantly reduced by using an empirical
relationship between the light-curve shape and the raw observed
(intrinsic) peak magnitude.  Different methods have been used to
perform this correction and we are now concerned with selecting a
single methodology for our combined sample.

\citet{1993ApJ...413L.105P} showed that the absolute magnitudes of
SNe\,Ia are tightly correlated with the initial decline rate of the
light-curve in the rest-frame $B$-band. They used a linear
relationship to correct the observed SN peak magnitudes, parameterized
via \m15, the decline in $B$-band magnitudes over the first 15 days
after maximum light -- intrinsically brighter SNe have wider, more
slowly declining light-curves. Further refinements by
\citet{1995AJ....109....1H,1996AJ....112.2391H} can reduce the
observed scatter to $\la0.17\,\rmn{mag}$, and to 0.11\,mag if
reddening corrections based on the SNe colours are made
\citep{1999AJ....118.1766P}.

An alternative technique, the Multi-color Light Curve Shape (MLCS),
introduced by \citet{1995ApJ...438L..17R,1996ApJ...473...88R}, adds or
subtracts a scaled `correction template' to a standard light-curve
template which creates a family of broader and narrower light-curves.
A simple linear (or non-linear) relationship between the amount of the
correction template added and the absolute magnitude of the SN can be
used to correct the SN magnitude, and results in a similar smaller
dispersion in the final peak magnitudes, particularly if multi-colour
light-curves are utilised in the fitting procedure.

A third approach, and the one used in this paper, is to parameterize
the SN light-curve time-scale via a simple stretch factor, $s$, which
linearly stretches or contracts the time axis of a template SN light
curve around the time of maximum light to best-fit the observed light
curves of every SN (Perlmutter et al. 1996, 97, 99). A SN's
`corrected' peak magnitude ($m^{\rmn{corr}}_{B}$) is then related to
its `raw' peak magnitude ($m^{\rmn{raw}}_{B}$) via

\begin{equation}
  \label{eq:1}
  m^{\rmn{corr}}_{B}=m^{\rmn{raw}}_{B}+\alpha(s-1)
\end{equation}

\noindent
where $s$ is the stretch of the light-curve time-scale and $\alpha$
relates $s$ to the size of the magnitude correction. As discussed in
\citet{2001ApJ...558..359G}, the stretch-correction technique is at
least as good at fitting existing data as any other single parameter
method can be.  It is important to note that, as we will show in later
sections, not applying any correction to the peak magnitudes makes no
significant difference to our conclusions.

Each SN (at both high and low redshift) was fitted to determine the
stretch factor $s$, using the published light-curves of the
low-redshift samples \citep{1996AJ....112.2408H,1999AJ....117..707R}
and the `exponential' SCP light-curve template used in P99.  The new
fits for the CfA SNe are very similar to those of the C-T SNe.  Each
SN peak magnitude is corrected to the value appropriate for a fiducial
light-curve corresponding to that of the template SN with $s=1$.
Finally, every stretch-corrected SN $B$-band peak magnitude is also
corrected for galactic extinction using the colour excess at the
Galactic coordinates of each SN \citep*{1998ApJ...500..525S}.

\section{New Data}
\label{data_section}

In this section we introduce the new dataset we have assembled for the
host galaxies of previously-discovered SCP high redshift SNe\,Ia.
Comparable data on the lower redshift SNe are taken from the articles
referenced in Section~\ref{sec:supernova-samples}.

The new data have been gathered through three observational programs:
(i) \hst\ observations which provide
spatially-resolved imaging data for morphologies, (ii)
high-resolution ground-based spectroscopy which characterises the
star-formation properties, and (iii) a variety of ground-based imaging
and low-resolution spectroscopic data which serves to add to our
capabilities of classifying the host galaxies. A component of the
latter survey includes the original SCP ground-based photometry of the
host galaxies drawn from the reference frames (Aldering et al., in
prep.).

\subsection{\textit{HST} Imaging}
\label{sec:hst-imaging}

\begin{figure*}
  \centering \includegraphics[angle=90,width=140mm]{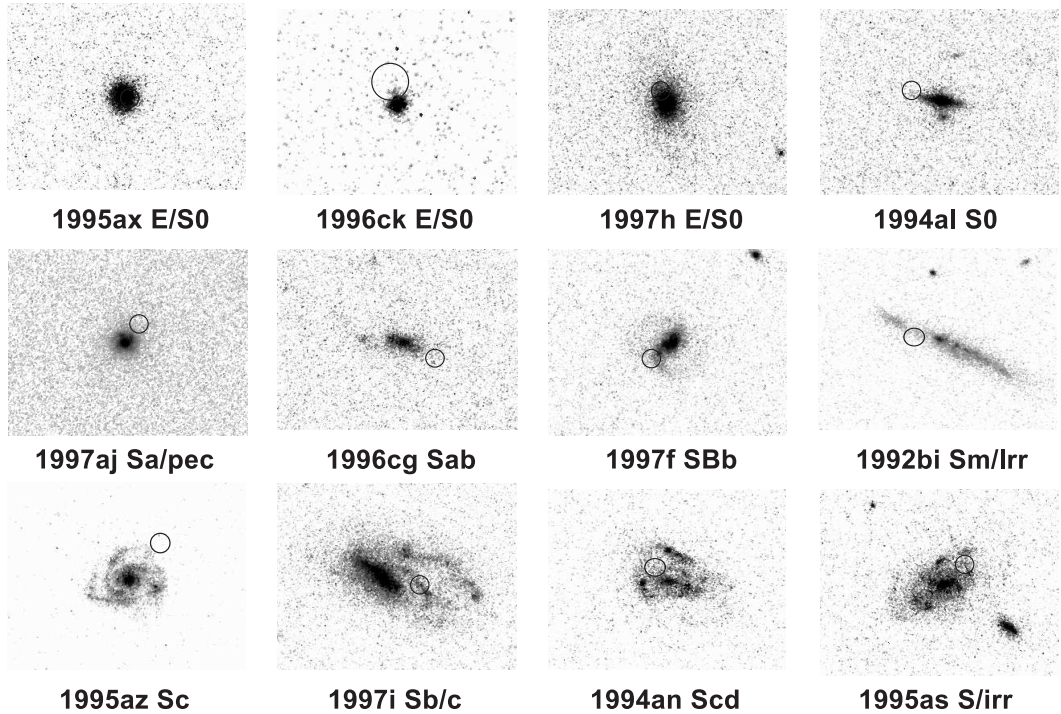}
\caption{
  A selection of the SNe\,Ia host galaxies imaged using STIS on the
  \hst, demonstrating the range of galaxy types and the quality of
  data produced by the program. In each image, the circle indicates
  the location of the (faded) SN derived via comparisons with existing
  ground-based data.}
\label{fig:hst_morphs}
\end{figure*}

Our \hst\ imaging data comes from two sources. The primary source is a
dedicated Space Telescope Imaging Spectrograph (STIS) snapshot program
(ID: 8313, 9131; PI: Ellis) of the host galaxies obtained during
cycles 8 and 10. The second source of \hst\ data arises via several
(some serendipitous) Wide Field Planetary Camera 2 (WFPC-2)
observations of a small number of the galaxies. A complete listing of
the host galaxies observed with \hst\ is given in
Table~\ref{tab:hstlog}. Examples of the STIS images can be found in
Fig.~\ref{fig:hst_morphs}.

\begin{table*}
\centering
\caption{\hst\ observing log for the host galaxies of the P99 high-redshift SNe discovered via the SCP collaboration.\label{tab:hstlog}}
\begin{tabular}{lccccl}
\small
SN Name & Program ID & Configuration & Exposure time (s) & Type$^a$ & Notes\\
\hline
1992bi &    8313 & STIS/50CCD  & $3\times434$ & 2 & Sm/Irr\\
1994al &    8313 & STIS/50CCD  & $3\times434$ & 0 & S0+comp\\
1994h  &    7786 & WFPC-2/606W  & $1\times100,1\times40$ & 1 & S/comp (see P97)\\
1994an &    8313 & STIS/50CCD  & $3\times434$ & 2 & Scd\\
1994am &    5378 & WFPC-2/814W  & $3\times2100$ & 0 & E (see P97)\\
(1994am &   9131 & STIS/50CCD  & $3\times434$ & 0 & E) \\
1994g  &    8313 & STIS/50CCD  & $3\times434$ & 2 & Pec/Irr\\
1994f  &    9131 & STIS/50CCD  & - & - &  Guide star acquisition failed\\
1995aq &    9131 & STIS/50CCD  & - & - & No observations\\
1995ar &    8313 & STIS/50CCD  & $3\times434$ & 1 & Sa\\
1995az &    8313 & STIS/50CCD  & $3\times434$ & 1 & Sc\\
1995ay &    9131 & STIS/50CCD  & - & - & No observations\\
1995at &    8313 & STIS/50CCD  & $3\times434$ & 1 & Sa\\
1995ba &    9131 & STIS/50CCD  & $3\times434$ & 0 & E/S0\\
1995ax &    8313 & STIS/50CCD  & $3\times434$ & 0 & E/S0\\
1995aw &    8313 & STIS/50CCD  & $3\times434$ & - & Low S/N\\
1995as &    8313 & STIS/50CCD  & $3\times434$ & 2 & S/Irr\\
1996cg &    8313 & STIS/50CCD  & $3\times434$ & 1 & Sab\\
1996cf &    8313 & STIS/50CCD  & $3\times434$ & 2 & Scd\\
1996ci &    9131 & STIS/50CCD  & $3\times434$ & 2 & Scd\\
1996cm &    9131 & STIS/50CCD  & $3\times434$ & - & Low S/N\\
1996cl &    5987 & WFPC-2/814W  & $6\times2600$ & 0 & S0 (see P99)\\
(1996cl)$^b$ &  (7372) & (WFPC-2/814W)  & ($6\times1100$) & (0) & (S0)\\
1996ck &    8313 & STIS/50CCD  & $3\times434$ & 0 & E/S0\\
1996cn &    8313 & STIS/50CCD  & $3\times434$ & 0 & S0\\
1997i  &    8313 & STIS/50CCD  & $3\times434$ & 1 & Sbc\\
1997h  &    8313 & STIS/50CCD  & $3\times434$ & 0 & E\\
1997g$^c$  &    9131 & STIS/50CCD  & $3\times434$ & 2 & Merging pair\\
1997f  &    8313 & STIS/50CCD  & $3\times434$ & 1 & SBb\\
1997j  &    9131 & STIS/50CCD  & $1\times434$ & 0 & E/S0;Interrupted exposure\\
1997l  &    8313 & STIS/50CCD  & $3\times434$ & 2 & Irr\\
1997o  &    9131 & STIS/50CCD  & - & - &  Guide star acquisition failed\\
1997p  &    9131 & STIS/50CCD  & - & - & No observations\\
1997s  &    9131 & STIS/50CCD  & - & - & No observations\\
1997n  &    9131 & STIS/50CCD  & $3\times434$ & 2 & Irr; poss. merger\\
1997q  &    9131 & STIS/50CCD  & $3\times434$ & 0 & S0\\
1997k  &    8313 & STIS/50CCD  & $3\times434$ & 2 & Irr\\
1997r  &    9131 & STIS/50CCD  & $3\times434$ & 1 & Sab\\
1997ac &    8313 & STIS/50CCD  & $3\times434$ & 0 & E/comp;low S/N\\
1997aj &    8313 & STIS/50CCD  & $3\times434$ & 1 & Spec\\
1997ai &    9131 & STIS/50CCD  & $3\times434$ & - & Low S/N\\
1997af &    8313 & STIS/50CCD  & $3\times434$ & 2 & Irr\\
1997am &    9131 & STIS/50CCD  & $3\times434$ & 2 & Spec/Irr\\
1997ap &    7590 & WFPC-2/814W  & $3\times700$ & - & Host not visible \citep[see][]{1998Natur.391...51P}\\
\hline
\end{tabular}
\flushleft
Notes:\\
$^a$ where: `0' indicates E/S0 type, `1' early-type spirals, and `2' late-type spirals/irregulars\\
$^b$ 1996cl has been observed twice by \hst/WFPC-2, prior to and during the SN event \citep*[see e.g.][]{2002MNRAS.332...37G}\\
$^c$ 1997g was also observed as part of 8313; however, this observation failed during guide star acquisition.\\
\end{table*}

In the case of the STIS data, each targeted galaxy was imaged in the
50CCD (clear aperture) mode, which approximates a broad $V+R+I$
band-pass, in a $3\times434\,\rmn{s}$ dithered series. Basic image
processing (bias subtraction, dark subtraction and flat-fielding) was
performed using the standard STIS reduction pipeline from within
\iraf. Cosmic ray removal and final image construction was then
performed using a `variable pixel linear-reconstruction' technique
using the \textsc{dither-ii} and \textsc{drizzle} packages
\citep{1997SPIE.3164..120F}. The output pixel size from this process
was $0.025$ arcsec across, and a value $\rmn{pixfrac}=0.6$ was used.

As well as this dedicated STIS snapshot program, we use WFPC-2 data
for one host from another SCP \hst\ program (ID:7590, PI: Perlmutter),
and also searched the \hst\ archive for additional \hst\ observations
of the hosts, finding three cases independently observed by
\hst/WFPC-2 (see Table~\ref{tab:hstlog} and references in P97 and
P99). The WFPC-2 data sets were again combined using the
\textsc{dither-ii} package used to reduce the STIS data.

The location of the faded SNe cannot be predicted with certainty on
the \hst\ (in particular STIS) images due to the pointing
uncertainties generic in snapshot observations. In some cases where
more than one galaxy is located near the centre of the \hst\ pointing,
this could confuse the correct assignment of the host galaxy. To
accurately locate the position of the SNe on the \hst\ (STIS or
WFPC-2) images therefore requires careful comparisons with existing
ground-based data taken in good seeing conditions. We performed a
cross-correlation between the \hst\ images and the existing calibrated
ground-based images (cf. Aldering et al., in prep.). In these images,
the SNe are visible, and hence the exact position can be determined in
relation to surrounding objects.  For the STIS field-of-view of
approximately $52\arcsec\times52\arcsec$, typically $6-8$ compact
stellar-like objects were common to both the deep ground-based and the
STIS data.  We calculate the relationship between the \hst\ and
ground-based pixel coordinate systems using \textsc{geomap} and
related tasks within \iraf, and are then able to transform the pixel
position of the SNe from the ground-based coordinate systems to the
\hst\ frame of reference.

We estimate the uncertainties in the derived positions using a simple
`bootstrap' technique. We calculate the location of the SN using all
the common objects, and then perform the same calculation but omitting
one common object each time. The resultant variation in the calculated
location of the SN gives approximate uncertainties in the derived
location using all the common objects. The process typically provides
SN positions which are accurate to $\simeq0.15$\,--\,$0.4$ arcsec; the
uncertainty is dependent on the number of objects in common between
the two images and the seeing of the ground-based data.  This accuracy
is, in all cases, sufficiently precise to unambiguously identify the
host galaxy.

This identification technique also enables us to derive projected
radial distances from the SNe locations to the centre of their host
galaxies for a sub-sample of the SNe.  The use of spatially resolved
\hst\ images has the advantage of being able to precisely locate the
centre of each galaxy reliably, something which is more difficult when
using ground-based images. The uncertainty therefore derives entirely
from the accuracy to which we can locate the SN position. In 20 of the
36 SNe with morphological information from \hst\ images, we are able
to calculate the SN location to an accuracy of $<0.3$\,arcsec; these
data will be discussed further in a later article.

We convert these distances in arcseconds to projected distances in kpc
using the angular diameter--distance relation; where necessary in
these calculations we assume a $\om=0.28, \ol=0.72$ and
$H_0=70\,\rmn{km\,s^{-1}\,Mpc^{-1}}$ cosmology.

\subsection{Keck spectroscopy}
\label{sec:keck-spectroscopy}

The pre-existing `discovery' host galaxy spectra were taken during the
period when the SN light dominated that of the host galaxy.  The
principal goal of those spectroscopic campaigns was to determine a
precise type and redshift for the SN candidate. As galaxies typically
have narrow emission and absorption line features, their presence can
often be detected even at maximum SN light; however, the contamination
by the SN usually makes any detailed study of the host galaxy
properties very difficult. We have therefore performed a second round
of optical spectroscopy for a sub-sample of the host galaxies imaged
by \hst\ using the Echellette Spectrograph and Imager (ESI) on the
Keck-II 10\,m telescope \citep[details on ESI can be found
in][]{2002PASP..114..851S}.  The goal of this component is to
considerably improve the diagnostic spectroscopy for the host
galaxies.

ESI can operate in three observing modes (imaging, low dispersion and
echellette); each galaxy was observed in the echellette and imaging
configurations. The echellette mode has a continuous wavelength
coverage of $\simeq4000$\,--\,11000\,\AA, spread across ten orders
with a constant $11.4\,\rmn{km}\,\rmn{s}^{-1}\,\rmn{pixel}^{-1}$
dispersion. All spectroscopic observations were performed using a slit
of width 1 arcsec. The position angle was calculated from the \hst\ 
images to ensure the slit passes through both the location of the (now
faded) SN and the host galaxy centre. Table~\ref{tab:grndobslog}
contains a log of the observations.

\begin{table*}
\centering
\caption{The ground-based observing log for the new data on the host galaxies of the SCP high-redshift SNe.\label{tab:grndobslog}}
\begin{tabular}{lcccc}
SN Name & Spectroscopic     & Spectroscopic     &  Imaging     & Imaging data\\
        &  observation date & exposure time (s) &  observation date & source\\
\hline
1992bi & 2001 May 21 & $2\times1200$ & 2001 May 21 & ESI\\
1994g  & 2000 Feb 10 & $3\times1200$ & 2000 Feb 10 & ESI\\
1994al & 2000 Nov 24 & $2\times1800$ & 2000 Oct 31 & COSMIC\\
1994an & 2000 Nov 24 & $2\times1800$ & ...     & ... \\
1995ar & 2001 Nov 13 & $2\times1800$ & ...     & ... \\
1995az & 2001 Oct 20 & $2\times1500$ & 2000 Oct 31 & COSMIC\\
1995ax & 2000 Nov 22 & $2\times1800$ & 2000 Oct 31 & COSMIC\\
1996cg & 2000 Feb 10 & $3\times1200$ & 2000 Feb 11 & ESI \\
1997i  & 2000 Feb 10 & $2\times1200$ & 2000 Feb 10 & ESI \\
1997h  & 2001 Oct 23 & $2\times1500$ & ...     & ... \\
1997f  & 2000 Feb 11 & $3\times1200$ & 2000 Feb 10 & ESI\\
1997aj & 2001 May 22 & $2\times1200$ & 2001 May 22 & ESI\\
1997g  & ...     & ...       & 2000 Feb 10 & ESI\\
\hline
\end{tabular}
\end{table*}

The primary data reduction steps were performed using the \makee\
data reduction software package written by T.
Barlow\footnote{\makee\ is available from\\
http://spider.ipac.caltech.edu/staff/tab/makee/index.html}. This
package produces a bias subtracted, flat-fielded, wavelength
calibrated and sky subtracted, but not order-combined, spectrum of
each object by using a master trace of a bright star and shifting
this trace to the position of the object to be extracted. A master
bias frame and internal (quartz lamp) flat-field were created on
every night of observation and used for that nights data; the
wavelength calibrations were performed using HgNeXe and CuAr lamps
bracketing each science observation.

On occasion, the automatic profile locating routine in \makee\ failed
for some orders or for those host galaxies which were spatially
resolved. In these cases, the object profile location on the slit and
the profile width were determined by hand and the spectrum extracted
with these parameters.

In order to combine the ten orders into a single continuous spectrum,
we performed flux calibration of the \makee-extracted spectra using
standard \iraf\ routines using a flux standard observed on the night
of observation at a similar airmass to the science target; suitable
standards were drawn from the lists of \citet{1988ApJ...328..315M} and
\citet{1990ApJ...358..344M}. We also correct for the extinction
properties of the Mauna Kea summit atmosphere
\citep[e.g.][]{1987PASP...99..887K,1990PASP..102.1052K}. Though an
absolute flux calibration is not possible with a slit of finite width,
a relative flux calibration between spectral regions is more reliable.
The primary uncertainty here arises from variations in the seeing as a
function of wavelength, where the FWHM of an image varies as
$\lambda^{-1/5}$. We estimate the effect to be at the $\sim5$ per cent
level between H$\alpha$ and H$\beta$, the primary emission lines that
interest us in this analysis.

After flux calibration, each order was examined and the wavelength
range offering the best signal-to-noise (S/N) determined.  These
wavelength ranges were then median combined to form a final
flux-calibrated spectrum from each individual exposure, and finally
the individual exposures median combined to form a final output
spectrum for each object.

\begin{figure*}
\centering
\includegraphics[width=140mm,angle=270]{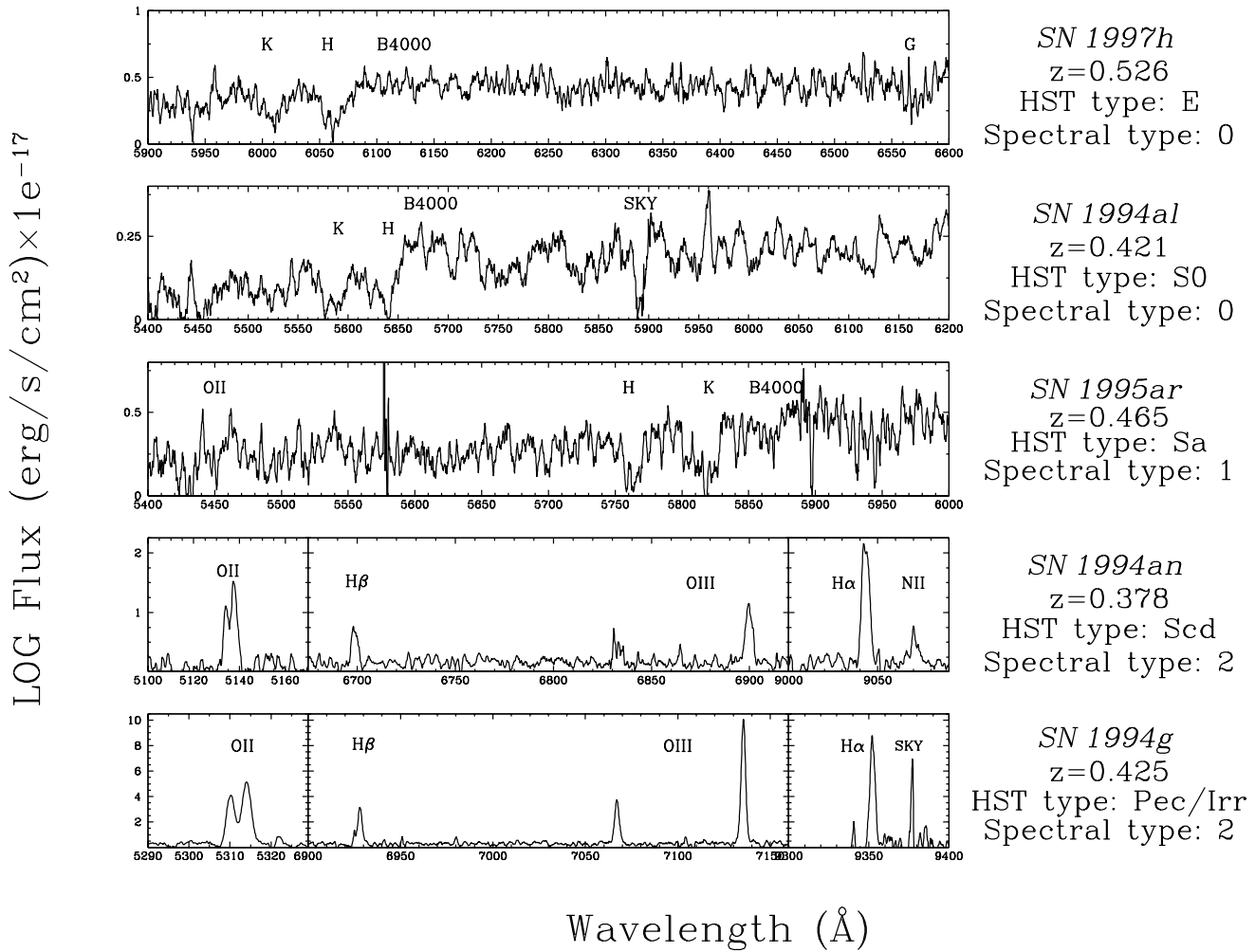}
  \caption{Example spectra taken from the ESI spectroscopic campaign demonstrating the range of spectral types and the quality of the ESI data obtained.}
  \label{fig:example_spectra}
\end{figure*}

The spectra were examined using the \splot\ facility in \iraf. For
each output spectrum a redshift was measured via visual inspection,
typically using Ca\,\textsc{ii} H,K absorption or \oii/Balmer
emission-line features. We find excellent agreement with the earlier
redshifts reported in P97/P99 in all cases.
Fig.~\ref{fig:example_spectra} demonstrates a range of the host
galaxy spectra observed; comments on each spectrum can be found in
Table~\ref{tab:spectra}. For the emission line spectra, we also
attempt to measure the strength of the \ha\ and \hb\ emission lines,
where they are present and are not contaminated by poor residual sky
subtraction from strong sky lines. These measures were also made using
\splot.

The \ha\ and \hb\ lines give an indication of the extinction in a host
galaxy, as the reddening nature of the dust makes the observed ratio
of the fluxes of two emission lines ($R_{\rmn{obs}}$) differ from the
intrinsic ratio ($R_{\rmn{int}}$). The colour excess of the ionised
gas, $E(B-V)_{\rmn{g}}$, can then be estimated via

\begin{equation}
  \label{eq:5}
  E(B-V)_{\rmn{g}}=\frac{\log(R_{\rmn{obs}}/R_{\rmn{int}})}{0.4[k(\lambda_{\hb})-k(\lambda_{\ha})]}
\end{equation}

\noindent
where $k(\lambda)$ is the form of the total-to-selective extinction of
the gas for a particular extinction law, and $\lambda_{\hb}$ and
$\lambda_{\ha}$ are the rest-frame emission wavelengths of \hb\ and
\ha\ respectively. Assuming case-B recombination, $R_{\rmn{int}}=2.87$
for $\ha/\hb$ \citep[e.g.][]{1989agna.book.....O}. The reddening
parameter $A_V$ can be calculated via $A_V=E(B-V)_{g}R_V$, where
$R_V\simeq3.1$ for a Milky Way-type extinction law
\citep[e.g.][]{1989ApJ...345..245C,1992ApJ...395..130P}. Where
available, these $A_V$ parameters are also listed in
Table~\ref{tab:spectra}; also listed are values where the extinction
due to the Milky-Way has been removed.  These $A_V$ parameters have
been corrected for stellar absorption affecting the \ha\ and \hb\ 
lines assuming stellar absorption equivalent widths of 2\AA\ in
absorption on each.  We make no corrections for the seeing variations
between the two wavelengths of interest, but note that as seeing
typically improves with wavelength (and hence slit losses will be
smaller), our measured \ha\ to \hb\ ratios will effectively be upper
limits on the intrinsic ratios.

One further subtlety is that these values of $A_V$ are measured for
the extinction on the nebular emission lines originating in
star-forming \hii\ regions. This may not, and indeed probably will
not, be representative for regions of the galaxy containing older
stellar populations where SN\,Ia may originate; indeed, it has been
shown that the continuum emission from stars in starburst galaxies is
often less obscured than the line emission from the gas
\citep[e.g.][]{1988ApJ...334..665F,1994ApJ...429..582C,1999A&A...349..765M}.
An estimate of this extinction can be derived using the prescription
of \citet{2000ApJ...533..682C}, who relate the colour excess of the
stellar continuum, $E(B-V)_{\rmn{s}}$, to $E(B-V)_{\rmn{g}}$ using
$E(B-V)_{\rmn{s}}=0.44\,E(B-V)_{\rmn{g}}$.

We discuss the utility of the spectroscopic data further in
Sections~\ref{sec:host-galaxy-class} and \ref{sec:discussion}.

\subsection{Ground-based Imaging}
\label{sec:ground-based-imaging}

Many of the original SN search images detect the host galaxy, and
photometric measures of these hosts are described and analysed in a
companion paper which includes the reference and late-time images free
from SN contamination (Aldering et al., in prep). As ESI offers an
direct imaging capability, we took the opportunity to gather
additional photometric measures when undertaking the spectroscopic
component described above. Additional photometry was obtained with the
COSMIC instrument at the Hale 200-in telescope at the Palomar
Observatory. An observing log of all the new imaging observations
obtained in addition to that described in Aldering et al. is given in
Table~\ref{tab:grndobslog}.

The galaxies were observed in filters which most closely matched those
filters used in the original SNe search program and in Aldering et
al., the standard Cousins $R$ and $I$ band filters (corresponding to
rest-frame bands $B$ and $V$ for most of our sample). In the case of
ESI we used standard $R$ and Gunn $i$ filters
\citep{1976PASP...88..543T}, and for COSMIC we used Gunn $r$ and $i$.
The new imaging data were reduced using standard techniques: overscan
subtraction, bias subtraction and flat-fielding were performed using
calibration data taken on the night of observation. The target host
galaxy was identified using the original SN detection images, and raw
magnitudes measured using the \sex\ program version 2.21
\citep{1996A&AS..117..393B}. We used the automatic aperture magnitudes
determined by \sex, which are similar to Kron's `first moment'
algorithm \citep{1980ApJS...43..305K}. As the $R$ and $I$ images were
taken consecutively on each night of observation, variations in the
seeing between the $R$ and $I$ band data are negligible.

All the imaging data were taken on photometric nights and standard
star fields \citep{1992AJ....104..340L,1994PASP..106..967J} were
observed at various airmasses to derive the CCD zeropoints and
variation of extinction with airmass in order to calibrate the
observed raw magnitudes on to the $\alpha$-Lyrae system. Finally, each
integrated magnitude was corrected for galactic extinction using the
colour excess maps of \citet{1998ApJ...500..525S}.

Seven of the 11 host galaxies imaged with ESI $R$-band filter
overlapped with the sample of Aldering et al., enabling us to
determine the photometric precision of the joint dataset. We directly
compare the $R$-band magnitudes derived from the two datasets, and
find an excellent agreement; the mean absolute difference between the
two datasets is only 0.06\,mag (in the sense that the ESI magnitudes
are fainter), with the largest discrepancy 0.12\,mag. This discrepancy
arises principally from the different magnitudes measured: Aldering et
al. define their magnitudes to include all the light from within two
Petrosian radii of a galaxy, whilst the definition of the magnitudes
in \sex\ leads to a slightly smaller aperture. These differing
definitions will not affect the calculated colours unless the galaxies
under study have marked colour gradients in their outermost regions.
In the future analyses in this paper where we use these colours to
help classify the hosts galaxies, we use data in the preference: i)
Keck-ESI $R$,$i$, ii) Cousins $R$,$I$ (Aldering et al, in prep.), and
iii) COSMIC $r$,$i$.

\section{Host Galaxy Classification}
\label{sec:host-galaxy-class}

We now turn to the question of characterising each host galaxy
according to the data presented above. In the case of the high
redshift SNe, we classified each host galaxy according to one of
three broad categories: spheroidal (E/S0, type 0), early-type
spiral (Sa through Sbc, type 1) and late-type spiral (Scd through
to irregular types, type 2), based on a combination of the three
diagnostics available to us.

In order of preference in the classification scheme, these are:

\begin{itemize}
\item \hst\ morphology. The morphologies of the galaxies were visually
  classified by two of the authors (RSE and MS). The morphology of each
  \hst-observed host is given in Table~\ref{tab:hstlog}; examples of
  each classification can be found in Fig.~\ref{fig:hst_morphs}.

\item Ground-based spectrum. The ESI spectrum (or if not available a
  pre-existing SN discovery spectrum with sufficient S/N) was
  classified according to the presence or absence of nebular emission
  lines and the prominence of the discontinuity around Ca\,\textsc{ii} H and K.
  Our scheme runs from 0 (no emission lines, strong break) to 2 (clear
  strong \hii-like spectrum, no break). Comments on each object can be
  found in Table~\ref{tab:spectra}; example spectra are shown in
  Fig.~\ref{fig:example_spectra}.
  
\item Ground-based integrated colour. We use a set of template SEDs
  \citep[taken from][]{1997A&AS..122..399P} to estimate the
  approximate galaxy class based on the observed $R$-$I$ colour and
  known redshift. The appropriate $R$ (or $r$) and $I$ (or $i$) filter
  response curves are convolved with each SED template at the host
  redshift, with the galaxy assigned a `colour--type' based on the
  nearest matching galaxy SED. We do not account for redshift
  evolution in the galaxy SEDs in this process.
\end{itemize}

\begin{table*}
\caption{ESI ground-based spectral classification\label{tab:spectra}}
\begin{tabular}{lcccl}
SN Host & Spectral type & $A_V$(raw)$^b$ & $A_V$(corrected)$^c$ & Comments$^a$\\
\hline
1992bi & 2 & 0.11 & 0.06 (0.03) & Broad \oii, \hb, \oiii \\
1994g  & 2 & 0.64 & 0.62 (0.27) & s. \oii, \hb, \oiii \\
1994al & 0 & -    & -    & s. H,K, B4000\\
1994an & 2 & 1.08 & 0.87 (0.38) & \oii, \hb, \oiii, \ha \\
1995ar & 1 & -    & -    & H,K, B4000, \oii, \hb\\
1995az & 1 & 3.71 & 3.16 (1.39) & \oii, \hb, \ha \\
1995ax & 0 & -    & -    & Flat; w. H,K, B4000\\
1996cg & 1 & -    & -    & w. \oii\\
1997i  & 1 & 1.15 & 1.01 (0.44)& s. \oii, \hb, \oiii, \ha \\
1997h  & 0 & -    & -    & s. H,K, B4000\\
1997f  & 1 & -    & -    & w. \oii, \hb, \oiii\\
1997aj & 1 & -    & -    & \oii, \hb, \oiii\\
\hline
\end{tabular}
\medskip
\flushleft
Notes:\\
$^a$B4000=4000\AA\ break, s.=strong, w.=weak\\
$^b$From Balmer decrement (corrected for stellar absorption), where available\\
$^c$Corrected for Galactic extinction, bracketed value is based on the \citet{2000ApJ...533..682C} prescription; see text.\\
\end{table*}

The question now arises as to how to combine these classifications,
recalling that not all diagnostics are available for each host galaxy.
We assigned a confidence of either 0 (probable) or 1 (secure), based
on the agreement and availability of the various measures. This leads
to two different samples of data -- a `securely' classified sample
(our primary analysis dataset) and a slightly enlarged, but less
securely classified, sample. Note that even in the less secure sample,
the discrimination between spheroidal and disc based systems is
usually robust, even if the precise class in the case of spiral
systems is not clear. The following scheme is used:

\begin{itemize}
\item The \hst\ imaging and ESI spectra; in all available cases these
  agree (Confidence `1'),
\item Clear \hst\ morphology; no ESI spectrum (Confidence `1'),
\item Ambiguous \hst\ morphology; supporting evidence from
  pre-existing (SN discovery run) spectral type or colour (Confidence
  `1'),
\end{itemize}

36 host galaxies (86 per cent of the total of 42 SNe in the P99
high-redshift sample) meet the above criteria. Three additional
galaxies have been less securely classified according to:

\begin{itemize}
\item No \hst\ image; class taken from good quality ground-based
  imaging \textit{and} spectroscopic data (Confidence `0').
\item No \hst\ image; class taken from good quality ground-based
  imaging \textit{or} spectroscopic data (Confidence `0').
\end{itemize}

\begin{figure*}
\centering
\includegraphics[width=120mm]{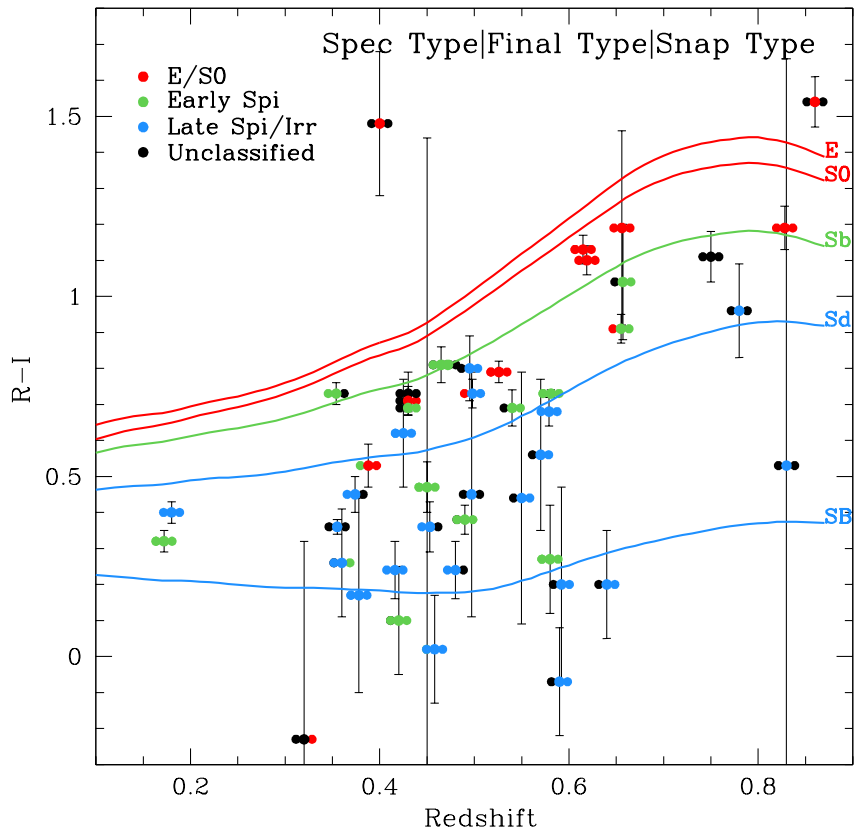}
\caption{
  The colour-redshift relation for the subset of SN host galaxies with
  both $R$ and $I$ measures, together with the classification of the
  SN host galaxies from the three diagnostics available (the colour
  type, \hst-type and spectral type). The colour shown is Galactic
  extinction corrected $R$-$I$; galaxies with colours measured based
  on Gunn filters are not shown (though similar diagrams exist for
  these smaller subsets). The final adopted classification is also
  shown. Overlaid SEDs are from \citet{1997A&AS..122..399P}, ranging
  from elliptical (E) to starburst (SB) galaxies. In the
  classification process, no redshift evolution in the galaxy
  properties is assumed to occur. Internal extinction in the host
  galaxies is not accounted for on this plot.}
\label{fig:zcol}
\end{figure*}

In total we therefore have 39 classified galaxies in the enlarged
sample (93 per cent of the total P99 high-redshift objects). The
classification statistics can be found in
Table~\ref{tab:classes_stats} and the agreement between the different
assigned types for the hosts can be examined in Fig.~\ref{fig:zcol}
in the context of the adopted ($R-I$) colour-redshift relation for
those galaxies for which $R$ and $I$-band measures are available.  The
classifications are colour coded from red (E/S0) through green (early
spiral) to blue (late spiral/irregular); black denotes those galaxies
with no assigned class; these data can also be found in
Table~\ref{tab:classes}.

\begin{table*}
\centering
\caption{The classification of the P99 42 high-redshift SN host galaxies\label{tab:classes}}
\begin{tabular}{lcccccccc}
SN name$^a$ & & Redshift & ESI ST$^b$& Other ST$^b$& Colour & \hst & Final & Classification\\
& &  & &  & Type & Type & Type & security\\
\hline
1992bi &   & 0.458 & 2 & 1 & 2 & 2 & 2 & 1 \\
1994al &   & 0.420 & 0 & 0 & - & 0 & 0 & 1 \\
 1994h & $\ast$ & 0.374 & - & 0 & - & 1 & 1 & 1 \\
1994an &   & 0.378 & 2 & 2 & 2 & 2 & 2 & 1 \\
1994am &   & 0.372 & - & 0 & - & 0 & 0 & 1 \\
 1994g &   & 0.425 & 2 & 2 & 1 & 2 & 2 & 1 \\
 1994f &   & 0.354 & - & 1 & 1 & - & 1 & 1 \\
1995aq &   & 0.453 & - & 2 & 2 & - & 2 & 1 \\
1995ar &   & 0.465 & 1 & - & 1 & 1 & 1 & 1 \\
1995az &   & 0.450 & 1 & 1 & 2 & 1 & 1 & 1 \\
1995ay &   & 0.480 & - & 2 & 2 & - & 2 & 1 \\
1995at &   & 0.655 & - & 0 & 1 & 1 & 1 & 1 \\
1995ba &   & 0.388 & - & 1 & 2 & 0 & 0 & 1 \\
1995ax &   & 0.615 & 0 & 0 & 1 & 0 & 0 & 1 \\
1995aw &   & 0.400 & - & - & 0 & - & 0 & 0 \\
1995as &   & 0.498 & - & 0 & 1 & 2 & 2 & 1 \\
1996cg & $\ast$ & 0.490 & 1 & - & 2 & 1 & 1 & 1 \\
1996cf &   & 0.570 & - & - & 2 & 2 & 2 & 1 \\
1996ci &   & 0.495 & - & - & 1 & 2 & 2 & 1 \\
1996cm &   & 0.450 & - & - & 2 & - & 2 & 0 \\
1996cl &   & 0.828 & - & 0 & 1 & 0 & 0 & 1 \\
1996ck &   & 0.656 & - & 0 & 1 & 0 & 0 & 1 \\
1996cn & $\ast$ & 0.430 & - & - & 1 & 1 & 1 & 1 \\
 1997i &   & 0.172 & 1 & 1 & 2 & 1 & 1 & 1 \\
 1997h &   & 0.526 & 0 & 0 & 1 & 0 & 0 & 1 \\
 1997g &   & 0.763 & - & - & 2 & 2 & 2 & 1 \\
 1997f &   & 0.580 & 1 & 1 & 2 & 1 & 1 & 1 \\
 1997j &   & 0.619 & - & 0 & 1 & 0 & 0 & 1 \\
 1997l &   & 0.550 & - & - & 2 & 2 & 2 & 1 \\
 1997o & $\ast$ & 0.374 & - & 2 & 2 & - & 2 & 1 \\
 1997p &   & 0.472 & - & 1 & 1 & - & 1 & 1 \\
 1997s &   & 0.612 & - & 2 & - & - & - & - \\
 1997n &   & 0.180 & - & 2 & 2 & 2 & 2 & 1 \\
 1997q &   & 0.430 & - & - & 1 & 0 & 0 & 1 \\
 1997k &   & 0.592 & - & - & 2 & 2 & 2 & 1 \\
 1997r &   & 0.657 & - & - & 1 & 1 & 1 & 1 \\
1997ac &   & 0.320 & - & - & 2 & 0 & - & - \\
1997aj &   & 0.581 & 1 & 1 & 2 & 1 & 1 & 1 \\
1997ai &   & 0.450 & - & - & - & - & - & - \\
1997af &   & 0.579 & - & 2 & 2 & 2 & 2 & 1 \\
1997am &   & 0.416 & - & 2 & 2 & 2 & 2 & 1 \\
1997ap &   & 0.830 & - & - & 2 & - & 2 & 0 \\
\hline
\end{tabular}
\medskip
\flushleft
Notes:\\
$^a$Starred SNe are excluded from the primary fit (`fit-C') of P99.\\
$^b$ST = Spectral Type.\\
\end{table*}

\begin{table*}
\caption{Statistics on the classification of the P99 42 high-redshift plus the extended low-redshift sample SN host galaxies\label{tab:classes_stats}}
\begin{tabular}{lrrrrr}
Host type & Low redshift$^a$ & \multicolumn{2}{c}{High redshift$^a$} & \multicolumn{2}{c}{Total$^{a}$}\\
 & & Secure & All & Secure & All\\
\hline
E/S0       & \textbf{ 3} ( 5) &  9 ( 9) & \textbf{10} (10) & 12 (14) & \textbf{13} (15)\\
Early Spi  & \textbf{14} (14) &  9 (12) & \textbf{ 9} (12) & 23 (26) & \textbf{23} (26)\\
Late types & \textbf{ 8} ( 8) & 14 (15) & \textbf{16} (17) & 22 (23) & \textbf{24} (25)\\
Small Distances & \textbf{14} (14) & - - & \textbf{19} (19) & - - & \textbf{33} (33)\\
Large Distances & \textbf{10} (12) & - - & \textbf{ 7} ( 7) & - - & \textbf{17} (19)\\
All        & \textbf{25} (27) & 32 (36) & \textbf{35} (39) & 57 (63) & \textbf{60} (66)\\
\hline
\end{tabular}
\medskip
\flushleft
Notes:\\
$^a$Bracketed numbers include SNe excluded from our  primary sample as stretch/residual outliers, or those likely reddened; see text and P99 for further details.\\
\end{table*}

As noted, the classifications of each object using the different
techniques agree well; in particular, there are no red objects with
emission line spectra or spiral morphology and no blue objects without
emission lines which have a spheroidal \hst-type.  The greatest
difficulty is in distinguishing between type 1 and type 2 objects,
where the $R-I$ colour is of only limited value, possibly due to the
increased presence of dust or evolutionary effects in these spiral
systems when compared to E/S0 objects; in these objects we mostly rely
on the other two diagnostics. Thus it is important to recognise the
most robust distinction in our survey lies between type 0 and the
remainder.

Classifications for the low-redshift SN host galaxy sample are
taken from \citet{2000AJ....120.1479H}. These authors give more
than adequate morphological classes for all of the SNe in the
low-redshift sample, and we simplify their system into the same
three broader classes that we use for the higher redshift sample.
Additional data on the projected distance (in arcsec) from the
host galaxy is taken from \citet{2000ApJ...542..588I}, which we
convert into a projected distance in kpc.

One potential concern with the consistency of the morphological typing
might be a mismatch between the filters used for the imaging at low
and high redshifts -- a morphological $k$-correction. However, the
breadth of the STIS filter in use should minimize this effect. The
low-redshift sample is typically typed using $B$ or $V$-band images
\citep[see][]{2000AJ....120.1479H}, which at the mean redshift of our
high-redshift sample would correspond to a $R$ or $I$ passband, which
falls within the $V+R+I$ response of the STIS 50CCD; indeed this was
one of the motivations for selecting the broad 50CCD observing mode.

\section{Results}
\label{sec:results}

With each galaxy classified according to its morphological or
spectral type, we now search for any environmental effects that
may be present in the combined sample of high and low redshift
SNe. This section presents our results in two broad categories.
Firstly, we examine the Hubble diagram and its associated
statistics as a function of the environment of the SN, both in
terms of the host galaxy morphology and galactocentric distance.
Secondly, we investigate possible correlations between the light
curve parameters of the SNe and its environment similarly
characterised. We discuss the broader implications of our findings
in Section~\ref{sec:discussion}.

\subsection{Hubble diagrams as a function of SN environment}
\label{sec:hubble-diagrams}

\begin{figure*}
\centering
\includegraphics[width=140mm]{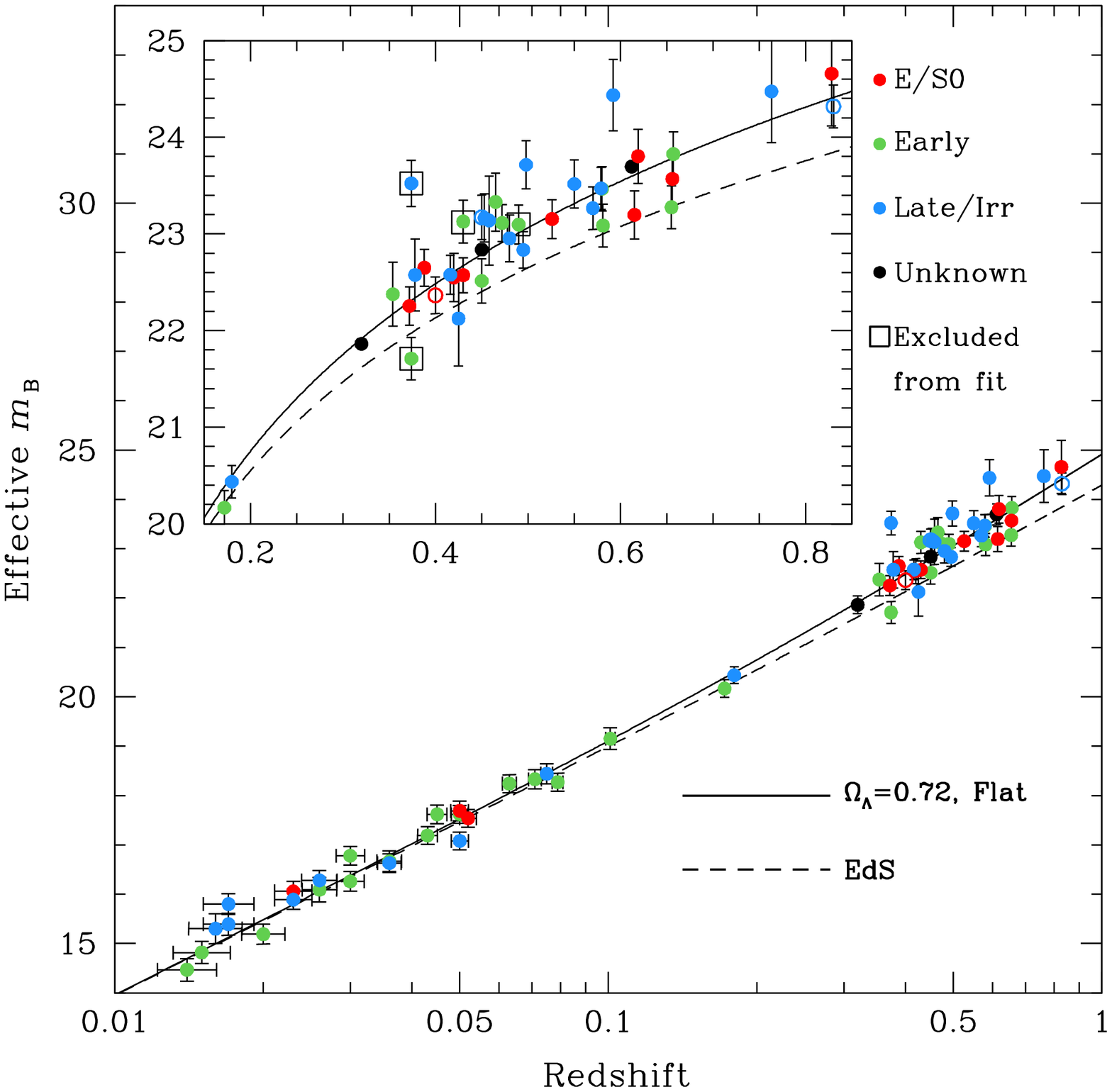}
\includegraphics[width=140mm]{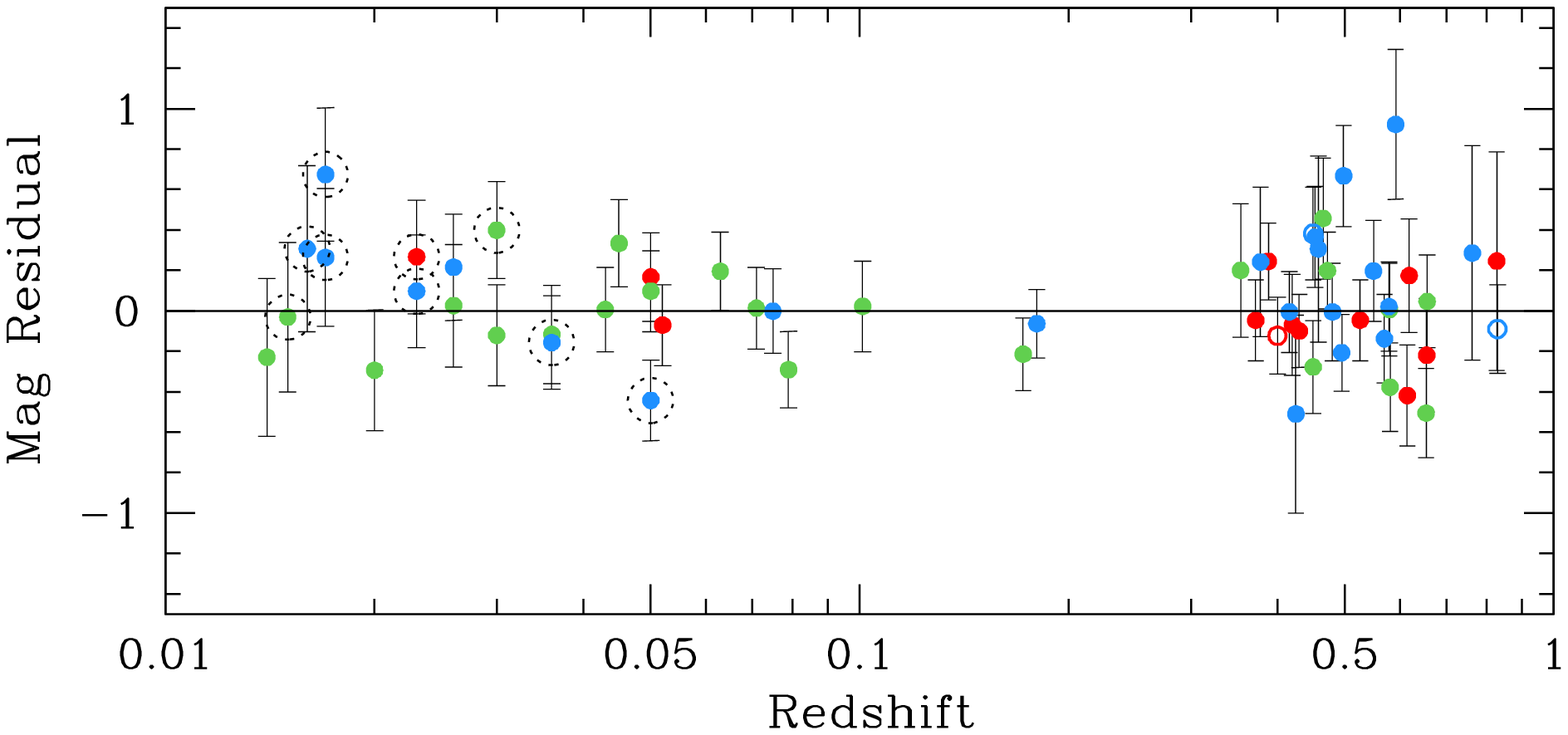}
  \caption{
    \textbf{UPPER PANEL: }The stretch-corrected SNe Ia Hubble diagram
    plotted according to the class of the host galaxy. The inset shows
    the high-redshift SNe, the main panel the Hubble diagram for the
    entire sample. Boxed points show high-redshift SNe excluded from
    the P99 solution. \textbf{LOWER PANEL: }The residuals from the
    adopted cosmology (`fit-C' of P99) for both high and low-redshift
    SNe.  Excluded SNe are not plotted. Dotted-circled points
    discriminate between CfA SNe and C-T SNe. In both panels unfilled
    circles denote galaxies with a less secure type-classification.}
  \label{fig:hubble_type}
\end{figure*}

We first examine the updated P99 Hubble diagram (fig.~2 in P99),
labelling each SN according to its host galaxy type. The upper panel
of Fig.~\ref{fig:hubble_type} shows this Hubble diagram in terms of
the effective rest-frame $B$ magnitude, i.e. corrected for the
light-curve stretch-luminosity relationship of equation~(\ref{eq:1}),
as a function of redshift for the 42 SCP high-redshift SNe, together
with the 25 SNe from the low-z C-T/CfA samples.  A zoom to the high
redshift sample is also shown. Those SNe excluded from the primary fit
of P99 are marked.

The SNe are colour-coded according to the host galaxy class.
Less-securely classified hosts are also marked. Those SNe with no
classes are marked unknown and ignored in the subsequent analyses.
The overlaid lines correspond to the apparent magnitude as a function
of redshift for a standard $s=1$ SN\,Ia expected in an EdS Universe
($\om=1$, $\ol=0$) and a flat, $\Lambda$-dominated Universe with the
best-fitting cosmology of P99 \citep[$\om=0.28$, $\ol=0.72$; see P99,
also][]{1998AJ....116.1009R}. The lower panel in
Fig.~\ref{fig:hubble_type} shows the residuals from this
best-fitting line for each SN.

We also perform additional fitting procedures for the SNe residing in
each galaxy class using a technique similar to that used in P99. We
consider two cases -- fitting for both \om\ and \ol\ independently,
and fitting for the case of a flat Universe (i.e.  fitting only for
\om\ or \ol\ and assuming $\om+\ol=1$).  This involves solving the
equation

\begin{equation}
  \label{eq:3}
  m_{B}^{\rmn{corr}}=m_{B}^{\rmn{raw}}+\alpha(s-1)={\cal M}_{B}+5\log D_L(z;\om,\ol),
\end{equation}

\noindent
where ${\cal D}_L\equiv H_0\,d_{\rmn{l}}$ is the `Hubble-constant
free' luminosity distance, and ${\cal M}_B\equiv M_B-5\log H_0+25$ is
the `Hubble-constant-free' $B$-band absolute peak magnitude of a
SN\,Ia with $s=1$ \citep[see e.g.][and P99]{1995ApJ...450...14G}.
This equation is solved for the parameters $\alpha$, ${\cal M}_B$,
\om\ and, for non-flat geometries, \ol, by least-squares i.e.
minimizing

  \begin{eqnarray}
  \label{eq:6}
    \lefteqn{\chi^2\left(\om,\ol,\alpha,{\cal M}_B\right)=}\nonumber\\ 
&& \sum^N_{i=1}\frac{\left(m_{Bi}^{\rmn{raw}} - \left[{\cal M}_B+5\log D_L\left(z_i\right)-\alpha\left(s_i-1\right)\right]\right)^2}{\sigma_i^2}.
  \end{eqnarray}

\noindent
As in P97 and P99, the small correlations between the high-redshift
SNe (due to shared calibration data) are accounted for using a
correlation matrix of the various uncertainties. We compare two
numerical minimization techniques: a simple Levenberg-Marquardt
technique, and a more complex Gauss-Newton method, and find an
excellent agreement between the derived parameters in both cases.

\begin{figure*}
\centering
\includegraphics[angle=270,width=180mm]{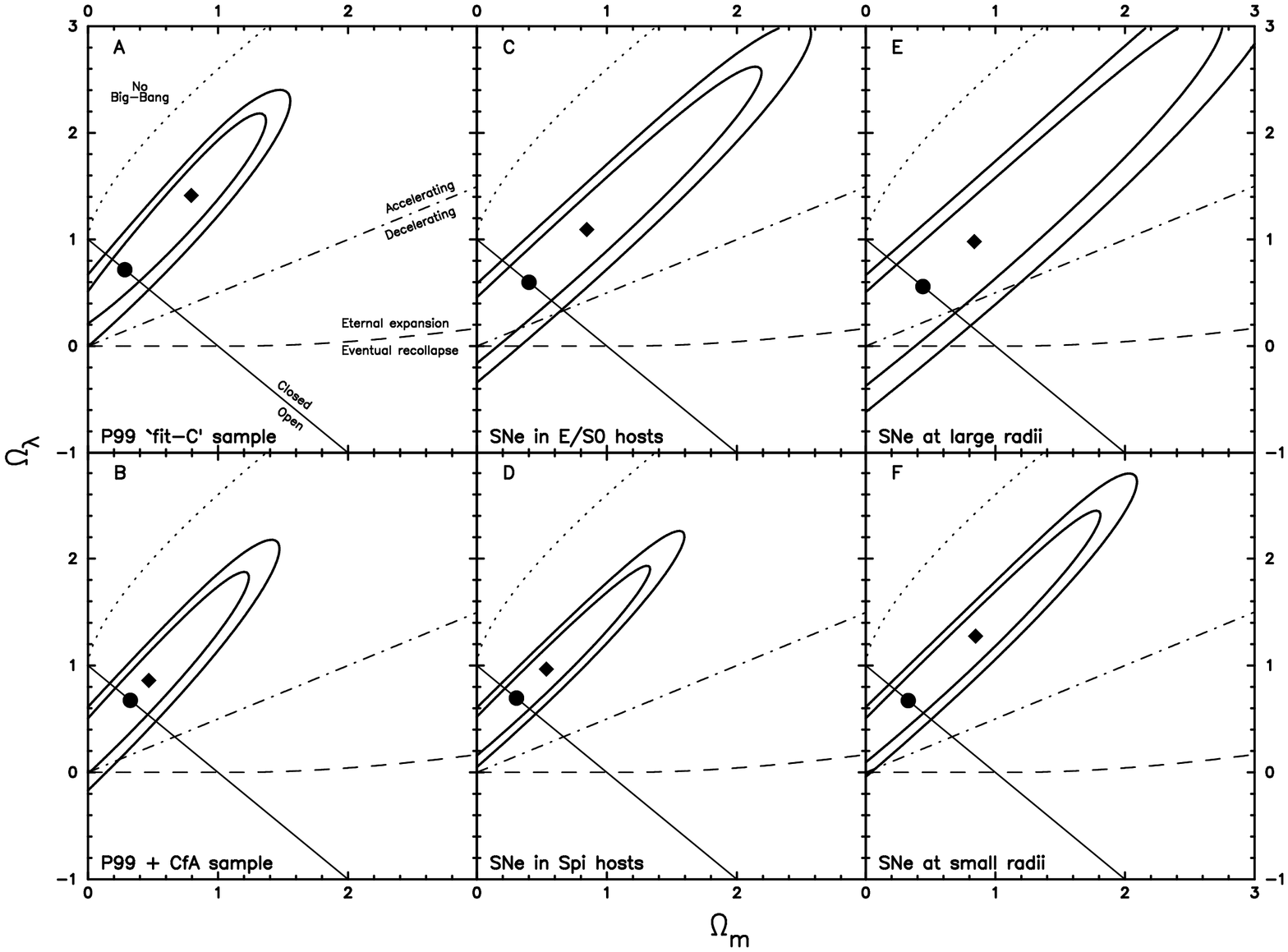}
  \caption{
    Confidence regions in (\om, \ol) for the fitting procedures in the
    different sub-samples listed in Table~\ref{tab:sne_cosmo_fits}.
    The ellipses correspond to 68 and 90\% confidence regions, and the
    best-fitting cosmologies are denoted by a filled diamond (general
    cosmology) and a filled circle (flat cosmology). The diagonal line
    corresponds to a flat Universe, the dotted line shows the boundary
    between `no big-bang' models, the dashed line denotes the
    infinite expansion boundary, and the dot-dashed line denotes those
    Universes which are currently undergoing acceleration. For those
    fits where $\alpha$ and ${\cal M}_B$ were fit (panels A and B),
    the confidence regions are obtained by integrating over these
    parameters (see text for details).}
  \label{fig:confidence}
\end{figure*}

Details of the best-fitting parameters are given in
Table~\ref{tab:sne_cosmo_fits}. The table also includes the fit
obtained for all of the SNe with a host galaxy type (regardless of
class) which provides a valuable consistency check that our `typed'
sample is representative of the larger one studied by P99. The
parameters $\alpha$ and ${\cal M}_B$ are determined only for the P99
SNe sample and for the new enlarged sample, and the parameters from
the latter fit adopted as fixed in the smaller sub-samples. (We study
variations in ${\cal M}_B$ in detail in Section~\ref{sec:discussion},
whilst the $\alpha$ parameter has only a small impact at high-redshift
due to the narrow stretch distribution seen, e.g. fig.~4 in P99).  For
these fits, the errors in the best-fitting $\alpha$ and ${\cal M}_B$
parameters are carried through in the analysis. For a flat $\ok=0$
Universe, Table~\ref{tab:sne_cosmo_fits} also lists the mean
dispersions from the best-fitting cosmologies for each sample, as well
as formal 1-$\sigma$ errors in \om\ derived from the co-variance
matrices of the various fits, assuming normally distributed errors.
The fitting procedure also generates two-dimensional confidence
regions, ${\cal P}(\om, \ol)\propto \exp(-\chi^2/2)$, shown in
Fig.~\ref{fig:confidence} for a selection of the SN sub-samples.  The
plot shows the 68\% and 90\% confidence ellipses in the (\om, \ol)
plane together with the best-fitting parameters for general and flat
cosmologies. For those datasets where $\alpha$ and ${\cal M}_B$ are
fit, as in P99 we calculate two-dimensional confidence regions by
integrating over these parameters, i.e., ${\cal P}(\om,\ol)=\int\int
{\cal P}(\om,\ol,{\cal M}_B,\alpha)\,\rmn{d}{\cal
  M_B}\,\rmn{d}\alpha$. This probability space also allows us to
estimate upper and lower error bounds on \om\ separately, which we
also report in Table~\ref{tab:sne_cosmo_fits}.

The probability distributions also allow us to calculate the
probability that $\ol>0$ and the probability that the Universe is
currently accelerating in its expansion for each SN sub-sample. The
latter can be expressed via the current, $z=0$, deceleration
parameter, $q_0$, where $q_0=\frac{\om}{2}-\ol$. As negative values of
$q_0$ indicate deceleration at the current epoch, the condition for
current acceleration in the expansion of the Universe is given by

\begin{equation}
\ol>\frac{\om}{2}.
\end{equation}

\noindent
The probabilities for these two classes of cosmological model are also
listed in Table~\ref{tab:sne_cosmo_fits}.

We also investigated potential environmental dependencies in the SN
properties by plotting the Hubble diagram with each SN labelled
according to its projected galactocentric distance
(Fig.~\ref{fig:hubble_dist}). SNe located at less then 7.5\,kpc
\citep[the approximate mean of the sample, as well as the value used
in the analysis of][]{1997ApJ...483L..29W} from the galaxy centre are
shown in red, those at larger distances are in blue. We also fit these
two additional subsets of SNe using the same procedure as above; the
results of these fits are again in Table~\ref{tab:sne_cosmo_fits}.

\begin{table*}
\centering
 \caption{The best-fitting cosmologies for SNe residing in each of the host galaxy types.\label{tab:sne_cosmo_fits}}
\begin{tabular}{lcccccccccc}
SNe Subset & Median & Best (\om,\ol)& Best \ol$^b$& Monte-Carlo & Mean & DOF$^d$ & $\chi^2$ & $\chi^2$ &\multicolumn{2}{c}{Probability that:}\\
 & redshift$^a$ & & & `error' & dispersion$^c$ &  &  &/DOF$^d$ & $\ol>0$ &$\ol\ge\frac{\om}{2}$\\
\hline
\multicolumn{11}{l}{\textbf{Fits with all parameters free:}}\\
P99 `fit-C' & 0.476 & 0.73, 1.32 & $0.72 ^{+0.09}_{-0.08}$ & $\pm0.07$ & 0.199\,(0.199) &  51 &  57 & 1.121 & 99.79 & 99.59 \\
P99 + CfA & 0.476 & 0.37, 0.67 & $0.65 ^{+0.11}_{-0.09}$ & $\pm0.06$ & 0.216\,(0.214) &  60 &  82 & 1.376 & 98.24 & 97.30 \\
Typed SNe & 0.480 & 0.46, 0.77 & $0.64\,(0.64)^{+0.11}_{-0.10}$ & $\pm0.07$ & 0.223\,(0.221) &  59 &  81 & 1.388 & 97.96 & 96.84 \\
 \hline
 \multicolumn{9}{l}{\textbf{Fits with $\alpha$ and ${\cal M}_B$ fixed:}}\\
E/S0 & 0.478 & 0.83, 1.02 & $0.57\,(0.58)^{+0.12}_{-0.11}$ & $\pm0.09$ & 0.159\,(0.173) &  12 &   8 & 0.680 & 97.09 & 94.80 \\
Spirals & 0.480 & 0.44, 0.81 & $0.67\,(0.66)^{+0.07}_{-0.06}$ & $\pm0.09$ & 0.235\,(0.234) &  46 &  73 & 1.590 & 99.85 & 99.75 \\
~~~Early & 0.472 & 0.62, 0.65 & $0.50\,(0.50)^{+0.12}_{-0.11}$ & $\pm0.14$ & 0.200\,(0.193) &  22 &  28 & 1.309 & 92.46 & 86.83 \\
~~~Late & 0.487 & 0.45, 1.03 & $0.77\,(0.78)^{+0.08}_{-0.07}$ & $\pm0.09$ & 0.272\,(0.274) &  23 &  40 & 1.761 & 99.98 & 99.97 \\
Large Distance & 0.458 & 0.72, 0.80 & $0.53 ^{+0.16}_{-0.14}$ & $\pm0.14$ & 0.196\,(0.204) &  15 &  19 & 1.276 & 94.38 & 89.46 \\
Small Distance & 0.498 & 0.77, 1.13 & $0.65 ^{+0.08}_{-0.08}$ & $\pm0.11$ & 0.256\,(0.251) &  32 &  56 & 1.777 & 99.57 & 99.27 \\

\hline
\end{tabular}
\medskip
\flushleft
Notes:\\
$^a$For high-redshift SNe\\
$^b$Assuming $\ok=0$; bracketed number is fit from securely classified sample\\
$^c$From best-fitting flat cosmology; bracketed value is the dispersion from P99 `fit-C' cosmology\\
$^d$For the $\ok=0$ fit.\\
\end{table*}

We checked the reliability of our quoted uncertainties in \ol\
(derived from the co-variance matrices of the various fits) by
performing detailed Monte-Carlo simulations of each SN data sub-set.
We created 10000 artificial datasets for each of the 9 sub-samples
(see Table~\ref{tab:sne_cosmo_fits}) by drawing $N$ SNe at a time
(with replacement) from the relevant sample, where $N$ is the number
of SNe in each set. We then fit these $N$ SNe in the same way as for
the original sample.  The mean of the resulting distribution in each
fitted parameter is satisfactorily close to the final fitted parameter
of the original dataset, with the standard deviation of the
distribution representing approximate 1-$\sigma$ uncertainties in the
various fits. These Monte-Carlo estimated uncertainties are also listed
in Table~\ref{tab:sne_cosmo_fits}.

We now turn to examining the significance of the various
correlations in the above plots. Fig.~\ref{fig:hubble_type} and
Table~\ref{tab:sne_cosmo_fits} reveal some interesting trends.

\begin{figure}
\centering
\includegraphics[width=80mm]{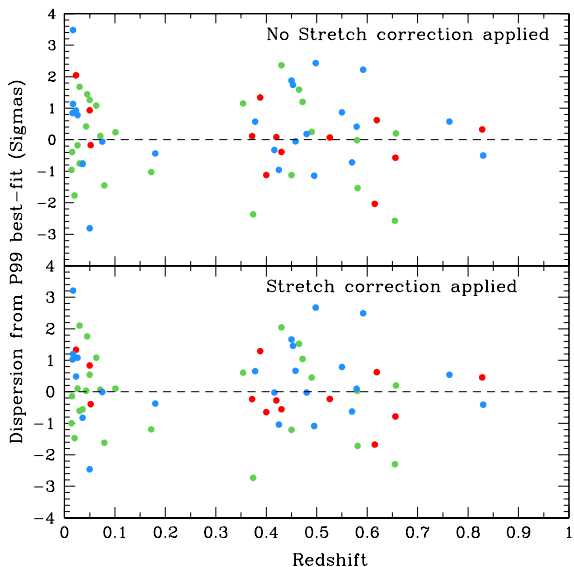}
  \caption{
    The dispersions (in sigmas) from the best-fitting cosmology of P99
    (`fit-C') as a function of host galaxy type, for both peak
    magnitudes corrected for the light-curve-width-luminosity relation
    (bottom) and uncorrected magnitudes (top). The colour coding is as
    in Fig.~\ref{fig:hubble_type}.}
  \label{fig:disp_comp}
\end{figure}

\begin{figure*}
\centering
\includegraphics[width=120mm]{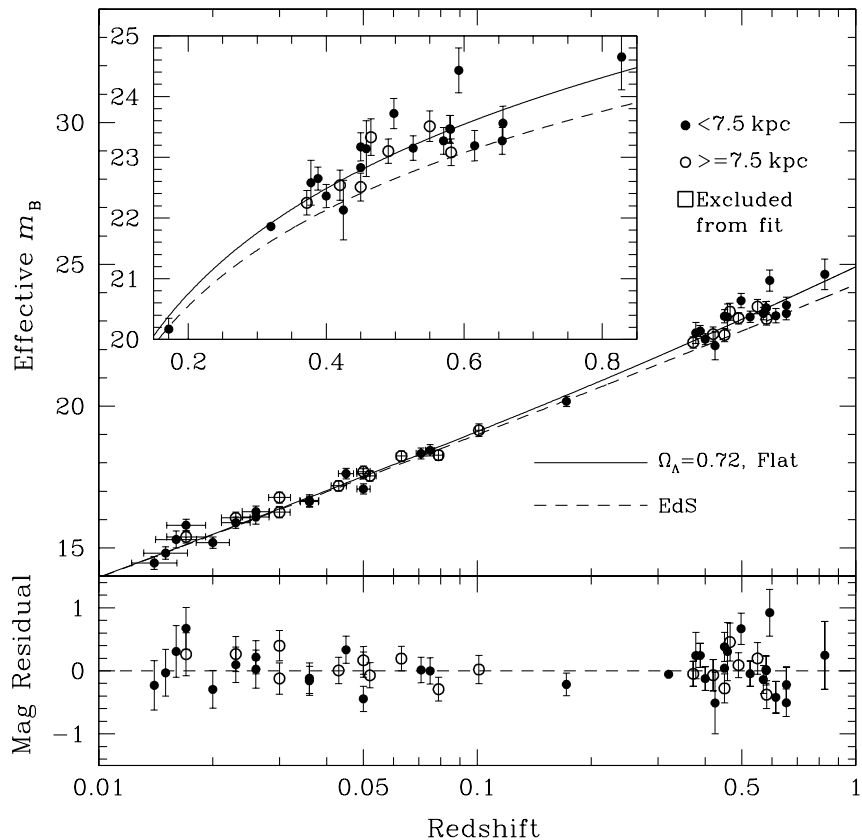}
  \caption{
    As Fig.~\ref{fig:hubble_type}, but with the SNe plotted according
    to the projected location from the centre of the host galaxy.
    Stretch corrections are applied. Filled circles show SNe lying at
    $<7.5\,\rmn{kpc}$, whilst empty circles show SNe at larger
    projected distances.}
  \label{fig:hubble_dist}
\end{figure*}

Firstly, each of the various samples and sub-samples strongly supports
the SCP/HZT conclusions for the presence of a cosmological constant;
indeed all sub-samples exclude an EdS Universe at greater than the 99
per cent confidence level and favour cosmological models with $\ol>0$
with $>97$ per cent confidence. If we restrict ourselves to flat
$\ok=0$ models, this confidence increases to $>99$ per cent for all SN
sub-samples. Most noticeably, the scatter about the best-fitting line
for SNe arising in spiral or irregular hosts is greater than that for
SNe arising in spheroidal host galaxies (see also the $\chi^2$/DOF in
Table~\ref{tab:sne_cosmo_fits}).  This trend of increased scatter with
class is particularly striking and can be seen more clearly by
examining the dispersions from the best-fitting cosmology
(Fig.~\ref{fig:disp_comp}).

If patchy dust extinction were the origin of the increased scatter in
the later type hosts, we would expect, on average, the SNe in these
systems to appear fainter. Such a trend is indeed seen; SNe occurring in
spiral galaxies appear fainter, with or without the application of a
stretch-correction. These conclusions are not affected if we only
consider the securely classified sample of host galaxies (see
Table~\ref{tab:sne_cosmo_fits}). We will return to the significance of
this conclusion in the next section.

The reduced scatter we observe for the E/S0 sample is an important
result. Accordingly, we are interested in understanding whether it
is a robust result. To verify this, we have performed simulations
to determine how likely we would be to obtain similar sub-samples
to the E/S0 sample by chance. We draw $N_{\rmn{ell}}$
high-redshift SNe from the full sample (where $N_{\rmn{ell}}$ is
the number of high-redshift SNe in E/S0 galaxies) and perform the same
best-fitting procedure to these SNe. By repeating the test a large
number (10000) of times, we investigated how frequently the derived
cosmological parameters produce a \ol\ smaller than the actual
E/S0 sample and a smaller mean dispersion about the best-fitting
cosmology.

We find that less than 3 per cent of the random samples possess a \ol\ 
smaller than that found for E/S0 galaxies, \textit{and} a mean
dispersion about the best-fitting cosmology smaller than that for the
E/S0 SNe.  Thus it seems the E/S0 sample is unlikely to be simply a
sub-sample selected randomly from the larger sample of SNe, and that
there is a significant difference in the photometric properties of the
E/S0 SNe that differs from the mean of the full SNe sample.

Finally, we can see no clear trends in the Hubble diagram plotted
according to the projected distance of the SN from the host galaxy
(Fig.~\ref{fig:hubble_dist}), again regardless of whether stretch
corrections are applied.

\subsection{Correlations between light-curve parameters and environment}

In studies of the environments of local SNe\,Ia, many authors
\citep[e.g.][]{1999AJ....117..707R,2000AJ....120.1479H} find that
intrinsically brighter SNe (SNe with larger stretch values) occurred
in late-type galaxies, and intrinsically fainter (small stretch) SNe
were preferentially located in spheroidal hosts.  It is important to
ensure that the effects we see in Fig.~\ref{fig:hubble_type} are not
systematic effects simply related to the application stretch
corrections which vary according to the type of the host galaxy
\citep[e.g.][]{2000AJ....120.1479H}.

We have considered this issue in two ways. First, we have repeated the
cosmological fits of Section~\ref{sec:hubble-diagrams}, comparing the
dispersions from the best-fitting cosmology with and without applying the
stretch corrections to each peak luminosity
(Fig.~\ref{fig:disp_comp}). We find that the trends in
Fig.~\ref{fig:hubble_type} remain even if stretch corrections are
not applied, with very similar derived cosmological parameters within
the uncertainties quoted in Table~\ref{tab:sne_cosmo_fits}.

\begin{figure}
\centering
\includegraphics[width=80mm]{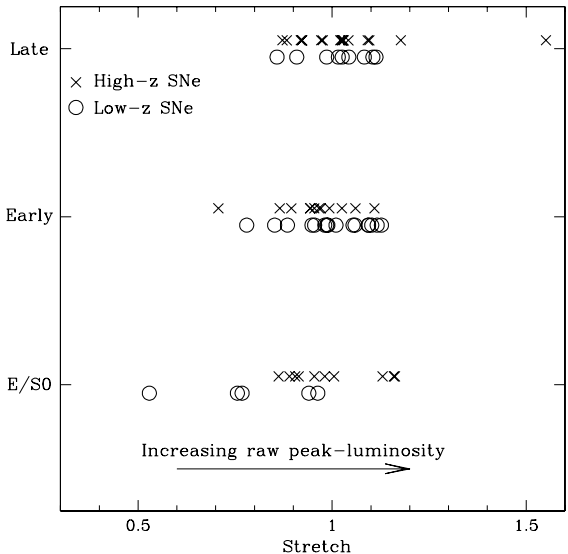}
  \caption{
    The distribution of SN stretch with host galaxy type. The
    low-redshift SNe are taken from the \citet{1996AJ....112.2408H}
    and \citet{1999AJ....117..707R} samples; see text for further
    details. All SNe (including those rejected as stretch outliers
    from the cosmological fits) are shown in this figure.}
  \label{fig:stretch_type}
\end{figure}

More directly, we can examine the distribution of the stretch
parameter for each SN located in each galaxy type.
Fig.~\ref{fig:stretch_type} shows this comparison for the high and
low redshift samples. The trend noted by \citet{1999AJ....117..707R}
and \citet{2000AJ....120.1479H} is apparent in the low-redshift
sample: an absence of small-stretch SNe in late-type galaxies and of
large-stretch SNe in early-type galaxies .  However, this same trend
is not seen in the SCP high-redshift SNe sample. This is due to the
fact that, as shown in P99 (their fig.~4), the stretch distribution of
the high-redshift SNe is narrower than that seen in the low redshift
samples. We return to this in Section~\ref{sec:discussion}.

\begin{figure}
\centering
\includegraphics[width=80mm]{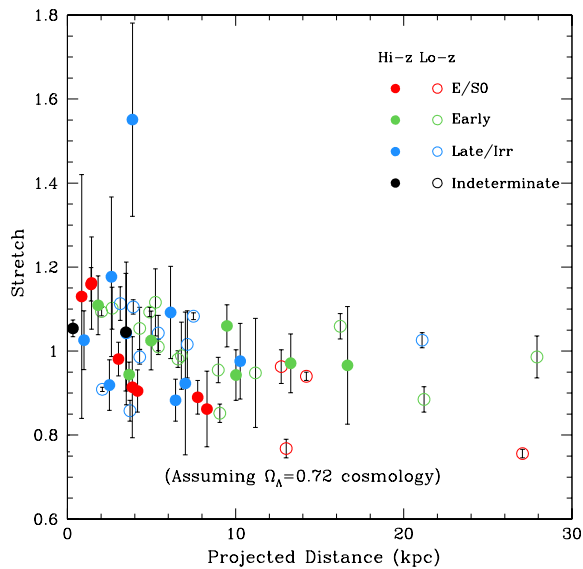}
  \caption{
    The SN stretch plotted as a function of projected distance from
    the centre of the host galaxy. The colour coding is as in
    Fig.~\ref{fig:hubble_type}.}
  \label{fig:stretch_dist}
\end{figure}

Finally, Fig.~\ref{fig:stretch_dist} shows the SN stretch plotted as
a function of projected distance from the host galaxy.  Here a weak
trend is found possibly suggesting that SNe at greater projected
distances may be fainter, similar to that seen in the local sample of
\citet{1999AJ....117..707R}.  This could possibly arise if, within a
distribution of luminosities, fainter SNe were systematically missed
close to the galaxy core \citep[e.g.][]{1979A&A....76..188S}, though
sensitivity calculations on the high-redshift sample suggest this is
unlikely to be the case
\citep{1996ApJ...473..356P,2002astro.ph..5476P}.

\section{Discussion}
\label{sec:discussion}

Of the various results presented in Section~\ref{sec:results}, the
most robust and significant trend we have found is that SNe found in
spiral and irregular galaxies present a larger scatter around the
best-fitting cosmological model than those located in E/S0 galaxies.
However, SNe in all galaxy types provide evidence for a significant
cosmological constant (at $\simeq5\sigma$). Additionally, SNe in
late-type galaxies also appear fainter than those found in E/S0s,
regardless of whether stretch-corrections are applied.

An obvious explanation of these main results, which do not depend on
whether stretch corrections are made, is that there is increased
extinction in later type galaxies and of an amount which varies
somewhat from galaxy to galaxy. Though the SNe magnitudes are
corrected for Galactic extinction, no account is taken of dust
residing either in the host galaxy or along the line of sight.

P99 investigated the effect of {\em differential reddening} on the
high-redshift SNe as compared to their local counterpart by
contrasting the mean $B-V$ colour of high redshift SNe at maximum
light with that for the local sample. They find that the colour excess
distributions show no significant differences at low and high
redshift, with the reddening distributions for the high-redshift
objects consistent with the reddening distribution for low-redshift
SNe within the measurement uncertainties, and both samples possessing
almost identical error-weighted means. Additionally, they perform
cosmological fits with each SN corrected for extinction according to
its \ebmv, and find that the fitted parameters change little with this
modification.

In the following we have revisited and extended this analysis.
Although the precision is poor for an individual SN, it is
advantageous to re-examine the situation as we now have independent
diagnostics which relate to the presence of host galaxy dust. The
differential reddening affecting any one SN can be determined by
comparing its observed \bmvmax\ with the $(\bmvmax)_{0}$ for
low-redshift SNe with the same stretch-factor.  The differential
colour excess \ebmv\ is determined as

\begin{equation}
  \label{eq:4}
  \ebmv=(\bmvmax)_{\rmn{obs}}-(\bmvmax)_{0}.
\end{equation}

The $(\bmvmax)_{\rmn{obs}}$ are calculated following P99. For SNe in
the redshift range $0.3<z<0.7$, the $R$ and $I$ band measures are used
instead of $B$ and $V$ and $(\bmvmax)_{\rmn{obs}}$ is calculated using
appropriate $k$-corrections
\citep*{1996PASP..108..190K,2002PASP..114..803N}. For SNe at $z>0.7$
the $R$ and $I$ measures correspond more closely to SN rest-frame $U$
and $B$ bands, so \ebmv\ is calculated from \eumb; likewise for SNe at
$z<0.3$, where the $R$ and $I$ correspond to rest-frame $V$ and $R$,
\ebmv\ is calculated from \evmr. We omit those 6 of the 42
high-redshift SNe that do not have both $R$ and $I$ measures from this
analysis.

\begin{figure}
\centering
\includegraphics[width=80mm]{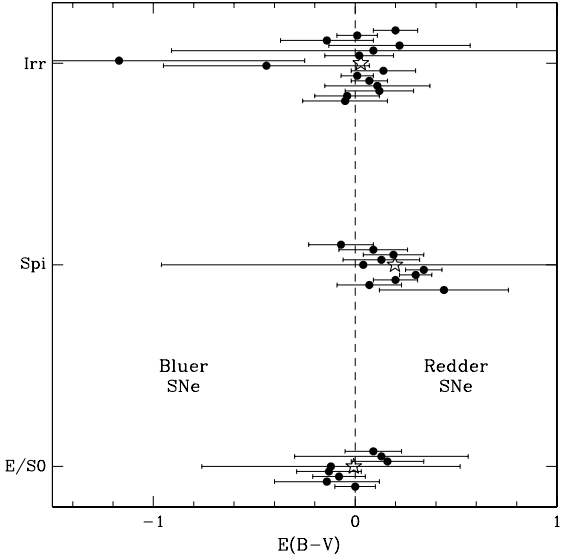}
  \caption{
    The SN rest-frame `colour excess' (\ebmv) plotted as a function
    of the host galaxy type. \ebmv\ is as defined in the text (see
    also P99). The error-weighted mean of each distribution are shown
    as stars; the error bars are derived from the $R$ and $I$
    photometry for each SN. The vertical offsets for each plotted SN
    is used to avoid over-plotting the data points and is illustrative
    only. Only the high-redshift SNe are shown.}
  \label{fig:colourexcess_type}
\end{figure}

The distribution of this \ebmv\ parameter as a function of the host
galaxy type is shown in Fig.~\ref{fig:colourexcess_type}.  The
errors in \ebmv\ are derived from the uncertainties in the $R$ and $I$
photometry. The error-weighted mean values of \ebmv\ for each class
are -0.01, 0.20 and 0.03 for E/S0, early spirals and later
spirals/irregulars respectively; the value for spirals taken as a
whole is 0.10.  We can see that SNe lying in E/S0 galaxies form a very
narrow and tight distribution around a mean $\ebmv\simeq0$, whereas
SNe residing in spiral galaxies show a larger range of \ebmv\ values
and means of $\ebmv>0$. This is suggestive that, as expected, SNe in
E/S0 galaxies are unlikely to be affected by the presence of dust,
whereas the effect of dust on SNe in spiral systems is more uncertain.
Furthermore, the two SNe rejected from `fit-C' in P99 as
`likely-reddened' (1996cg and 1996cn) are both located in early spiral
galaxies; indeed, these two objects are the primary cause of the
large mean value of \ebmv\ found for these systems.

The above result is, to some extent, independent of the morphological
trends seen in the Hubble diagram since the former relies on the SN
colour whereas the latter depends on the peak luminosity. Naturally
there are possible inter-dependencies but we argue both results
provide strong support for our two main conclusions. Firstly, much of
the scatter in the SNe\,Ia Hubble diagram arises from extinction in
the later type galaxies. As this scatter is host galaxy
morphology-dependent, clearly it is internal to the host galaxies.
Secondly, not only do we see a tight scatter in the Hubble diagram for
SNe in E/S0s, but differentially there is no obvious difference in the
extinction as inferred from the SNe at high and low redshift.
Accordingly, the cosmological conclusions derived from this E/S0
subset must be particularly robust to systematic errors arising from
extinction. As can be seen in Table~\ref{tab:sne_cosmo_fits}, this
data provides a $\simeq5\sigma$ detection of a non-zero cosmological
constant.

\begin{table*}
\centering
 \caption{The best-fitting ${\cal M}_B$ for SN\,Ia arising from different galaxy types, assuming the best-fitting cosmology for the E/S0 sub-sample.\label{tab:implieddust}}
\begin{tabular}{ccccc}
&E/S0 & Early Spi & Late Spi/Irr & All Spirals\\
\hline
${\cal M}_B$ & $-3.39\pm0.07$ & $-3.37\pm0.08$ & $-3.25\pm0.06$ & $-3.32\pm0.05$\\
\hline
Difference with E/S0 sample& -- & $0.02\pm0.11$ & $0.14\pm0.09$ & $0.07\pm0.09$ \\
\hline
\end{tabular}
\medskip
\flushleft
\end{table*}

An interesting question then is: \textit{how much extinction in
  high-redshift SNe\,Ia is implied by our results?} We can address
this question in the following manner. We repeat our cosmological fits
for the E/S0 and various spiral SNe sub-samples, but this time fit for
the parameter ${\cal M}_B$ rather than \om\ and \ol, and instead hold
\om\ and \ol\ fixed at their values as derived from the (presumed
dust-free) E/S0 sample of SNe. As ${\cal M}_B$ is the intrinsic
$B$-band peak magnitude of a SN\,Ia, if those SNe in E/S0 galaxies are
less affected by dust than those in spiral galaxies, we might expect
the best-fitting ${\cal M}_B$ to be brighter in these galaxies. These
best-fitting ${\cal M}_B$ values and 1-$\sigma$ errors are given in
Table~\ref{tab:implieddust}.

Assuming the best-fitting E/S0 flat-Universe cosmology, SNe appear
$0.14\pm0.09$\,mag fainter in late-type spirals than in E/S0s.
However, these uncertainties are currently large (due to the small
number of objects in the different samples), and the result is
consistent with both a near-zero offset and an offset of 0.23\,mag.
Though there are obvious caveats associated with this result, and the
scatter from SN to SN is large due to the photometric uncertainties in
the SN magnitudes, this measure would however appear to suggest that
the amount of extinction suffered by these high-redshift objects is
small, confirming earlier conclusions (e.g.  P99).

This result appears in contradiction to the work of
\citet{2002MNRAS.332..352R}, who suggest that the evidence for a
positive cosmological constant is reduced if an average internal
dust-extinction correction of 0.33\,mag for all high-redshift SNe is
made. However, median extinction values as large as this do not appear
consistent with these new results, particularly in the E/S0 systems
where internal extinction is expected to be very small, and which show
that $\ol>0$ at $\simeq5\sigma$.

Modest extinction for even the SNe occurring in late-type galaxies is
supported by the small amount of spectral data available for the
emission line host galaxies (Table~\ref{tab:spectra}), though the
extinctions implied are not as small as those indicated by the
dispersions in the Hubble diagram. For the 6 galaxies where a measure
of the Balmer decrement is possible (all at $z<0.5$ due to \ha\ 
shifted out of the ESI spectral coverage) the measures suggest (except
for one case) small extinctions on the nebular emission lines, and
hence smaller extinction on the stellar continuum likely affecting the
observed SNe, using the prescription of \citet{2000ApJ...533..682C}.
However, most of the measured $A_V$ are still larger than the
0.14\,mag implied by the difference in ${\cal M_B}$ between E/S0 and
spiral systems (Table~\ref{tab:implieddust}).  We hypothesise this is
due to SNe lying well away from the \hii\ regions from which the
nebular lines used to measure the dust content originate, an effect
which is not removed even after using the \citet{2000ApJ...533..682C}
prescription. In the one case where the extinction measured is large
($A_V\simeq3$, 1995az), the SN was located at a large projected
galactocentric distance of $\sim13\,\rmn{kpc}$, one of the largest in
the sample (see Fig.s~\ref{fig:hst_morphs}
and~\ref{fig:stretch_dist}), where the large amounts of dust that were
detected in the \hii\ regions may not be present. Clearly larger
samples of spectra are required to confirm these trends.

At first sight, the small extinction found for SNe in late-type
galaxies appears a surprising result. Observational evidence suggests
that the amount of star-formation in normal field galaxies increase
with redshift \citep[e.g][]{1996ApJ...460L...1L}, and that much of
this increase results from late-type and irregular systems
\citep{1998ApJ...499..112B}, which presumably should be dustier
systems than their non-starforming counterparts, as increased
star-formation activity is usually associated with increased dust
content.  Although we see such a correlation in our data, in absolute
terms the SNe of P99 even in the later-type hosts appear to suffer
little extinction, as confirmed by the limited spectroscopic data
presented here and the colours of the low and high-redshift SNe at maximum
light.

\citet{1998ApJ...502..177H} have investigated the effects of
extinction on the optical properties of SNe. For simple assumptions
concerning the spatial distribution of dust and SN progenitors in a
model disc galaxy, they find that the amount of extinction suffered by
SNe is small. The typical ranges in mean $A_B$ is 0.3\,mag to
$>1$\,mag when all SNe are considered; however these means are
dominated by a few high extinction events -- most SNe are mildly
obscured, but with a long tail to higher extinction events, which drop
out of flux-limited surveys such as the SCP/HZT search campaigns. Mean
values for an `extinction-limited subset' ($A_B<0.60$) range from 0.12
to 0.16\,mag, depending on the inclination of the system, in good
agreement with the 0.14\,mag difference we find between E/S0 and
late-type spiral SNe. 

A further explanation for the low extinction on high-redshift SNe in
late-type systems is that the progenitors of SNe\,Ia are largely
located in the outermost components of disc and irregular galaxies,
perhaps as landmarks of earlier stages of galactic assembly, whereas
the dust is distributed within regions of more recent star-formation
activity confined to the thin disc. In popular hierarchical models,
where the bulk of galactic assembly occurs over $1<z<3$, one would
expect much mixing of dust and old stars. The absence of significant
extinction in the bulk of our sample might be used to argue against a
significant increase in the merger rate prior to $z\simeq0.5-0.8$.

Another possible explanation is related to the selection effects
inherent in SN\,Ia surveys; do SN\,Ia searches sample the entire range
of galaxy types available at a given redshift? Alternatively, are we
missing SNe that reside in particularly dusty host galaxies? In
principle, we can examine this by comparing the morphological mix of
our host galaxy sample to the observed mix of normal galaxies in the
same redshift range.  The most relevant study to date is that of
\citet{1998ApJ...499..112B}, who analyse \hst\ imaging of the galaxies
selected in the CFRS and LDSS redshift surveys. They classify their
sample into three broad classes -- E/S0s, spirals and irregulars --
and find a large increase with redshift in the fractional number of
irregular objects from 9 per cent at $z<0.5$ to over 30 per cent at
$0.7<z<0.9$, with decreases in the relative numbers of E/S0s and
regular spirals. This leads to a change in the split of both
integrated $B$-light luminosity density
\citep[see][]{1998ApJ...499..112B}, \textit{and} a change in the
morphological split of integrated stellar mass density
\citep[see][]{2000ApJ...536L..77B}, both likely related to SN\,Ia
events. Hence, in a truly volume-limited survey of SNe\,Ia, we might
expect the fraction of events found in irregular galaxies to sharply
increase at $z>0.5$, tracing the increased stellar mass in these
systems.

Comparisons with this study are obviously problematic due to the small
size of the current sample and the differing redshift ranges, as well
as differing classification techniques for the galaxies. We have 16
securely classified galaxies at $0.5<z<0.8$, of which 7 are classed as
type-2 ($44\pm17$ per cent). However, the Brinchmann et al. scheme
would class many of these regular late-type galaxies separately from
those which are irregular systems -- only 3 of the objects ($19\pm11$
per cent) are irregular in this definition scheme. Brinchmann et al.
find a fraction of $33\pm6$ per cent, opening the possibility that we
do not see this same rapid rise in the number of `truly' irregular
systems in the SN sample.

This possible lack of irregular systems can be reconciled if we
hypothesise that such objects contain, on average, larger amounts of
dust due to their presumably higher star-formation activity. This
would result in any SNe located in these galaxies to be dimmed to a
greater extent than those in E/S0 or regular spiral systems; the SNe
searches could therefore be (implicitly) biased against extincted SNe.
Conceivably, many SNe\,Ia {\em do} suffer significant extinction but,
because of the flux-limited SNe search criteria, they are not included
in our sample.

Our observation that SNe\,Ia residing in E/S0 galaxies show a smaller
scatter about the best-fitting cosmological model (as revealed by the
$\chi^2$/DOF for the various fits in Table~\ref{tab:sne_cosmo_fits})
allows us to estimate the intrinsic spread in the peak magnitude of
these SNe compared to those residing in other classes of galaxy.
Currently, each SNe error contains a contribution of 0.17\,mag added
in quadrature to allow for the uncertainty in the
light-curve-width-luminosity correction
\citep[e.g.][]{1996AJ....112.2391H}, regardless of the host class. As
some of this dispersion arises from host galaxy extinction, we can use
the (presumed) dust-free E/S0 SNe population to estimate the intrinsic
dispersion for this class of SNe. For SNe in each host type, we
therefore repeat the cosmological fits, altering the intrinsic
dispersion until the $\chi^2$/DOF for each fit is equal to one. We
perform this test for the high and low-redshift SNe taken together,
and also solely for the high-redshift SNe.

The intrinsic dispersions are (in magnitudes) 0.09(0.10), 0.29(0.25),
and 0.69(0.24) for E/S0, early and late-type spirals/irregulars
respectively -- bracketed values refer to fits where only
high-redshift SNe are used (the values for spirals taken as a whole
are 0.46 and 0.26) . Note that the small intrinsic dispersion seen for
those SNe in E/S0 matches very closely the dispersion seen in
low-redshift SNe after a technique correcting for the effects of
reddening has been applied \citep[0.11\,mag;
see][]{1999AJ....118.1766P}, and the large dispersion seen in
late-type spiral classes is heavily influenced by one or two
low-redshift outliers. The implication from this small dataset is
clear: SNe residing in E/S0 hosts are superior `standard candles' than
those in later-type galaxies due to the small amount of extinction
affecting them.

Finally, it is important not to overlook a possibly important result
suggested by the lack of range in light-curve width (or stretch)
indicated in Fig.~\ref{fig:stretch_type}. This figure does provide
some evidence of a shift in the population distribution between SNe
found at low and high-redshift: at high-redshift we see no trend in
stretch with galaxy type, whereas at low-redshift SNe with larger
stretches are found in later-type galaxies and are missing from E/S0
galaxies.  It's important to emphasise that this population shift has
no impact on the security of the cosmological results, as
light-curve-width differences are accounted for in the stretch
correction of equation~(\ref{eq:1}), and stretch outliers are in any
case rejected from the fits. Given the small number of SNe in each
group, Fig.~\ref{fig:stretch_type} will, of course, be sensitive to
one or two extreme or unusual events.  Bearing this caveat in mind, we
performed Kolmogorov--Smirnov (K--S) tests on the different stretch
distributions for SNe residing in each galaxy class, and find the
significance levels for the hypothesis that the SNe at low and
high-redshifts are drawn from different distributions to be 89, 71 and
7 per cent for E/S0, early- and late-type galaxies respectively (the
significance level for low-redshift SNe in E/S0 and late-type galaxies
having a different stretch distribution is 94 per cent in this
sample). Though these trends are only suggestive given the small
number of events, they might be explained if we assume that the
light-curve properties of SNe\,Ia depend primarily on the \textit{age}
of the progenitor system.

At low redshift, it appears intrinsically fainter SNe are located in
older (E/S0) stellar environments rather than young spiral systems
\citep[e.g.][]{2000AJ....120.1479H}. As we look to higher-redshift
(i.e. viewing galaxies at an earlier stage in the evolution of their
composite stellar populations), the E/S0 galaxies will posses younger
populations than is the case in the local Universe (at $z=0.5$, the
age of the Universe is $\sim8\,\rmn{Gyr}$ compared to
$\sim13\,\rmn{Gyr}$ at $z=0$ in a $\Lambda$CDM cosmology). If, as was
discussed in Section~\ref{sec:introduction}, older progenitors
generate fainter SNe events, at high-redshift we would not expect to
see the fainter (smaller stretch) SNe found in local, older, E/S0
galaxies \citep{2000AJ....120.1479H}, thus strengthening the case for
a cosmological constant.

\section{Conclusions}
\label{sec:conclusions}

We have investigated the Hubble diagram of Type Ia supernovae spanning
a large range in redshift and have classified 39 distant events
according to the nature of their host galaxies as revealed by \hst\ 
imaging, intermediate dispersion spectroscopy and broad band colour.
Together with morphological data for the host galaxies of 25 local SNe
drawn from the literature, we find the following.

\begin{enumerate}

\item The scatter on the SN\,Ia Hubble diagram, as determined from
  residuals of the peak brightness with respect to the best fit
  cosmological model, correlates closely with the host galaxy type.
  The scatter is minimal for SNe occurring in host galaxies broadly
  classed as early-type, and increases towards later classes. The
  correlation does not depend on whether the SN are corrected for
  differences in light-curve shape and thus cannot arise primarily
  from type-dependent light-curves.
  
\item The above trend is accompanied by the observation that SNe
  occurring in the later types are \textit{on average} marginally
  fainter than those in E/S0 galaxies, with a difference in their
  absolute peak luminosities of $0.14\pm0.09$\,mag.
  
\item Following the approach of \citet{1999ApJ...517..565P}, we
  compare the colour-excesses of distant SNe\,Ia at maximum light with
  those observed locally for examples of the same stretch and see an
  indication of a type-dependent colour excess of a similar nature,
  with, on average, spirals hosting redder SNe.

\end{enumerate}

The data are also suggestive of a lack of any significant correlations
at high-redshift between the light-curve stretch parameter, which
correlates with the \textit{observed} peak luminosity, and the host
galaxy type. The full implications of this result are currently
unclear, given the (current) small sample sizes.  However, the lack of
small stretch (fainter) SNe and the increase in larger-stretch SNe in
high-redshift E/S0s, compared to that found in low-redshift samples,
suggests that the age of the progenitor system may be the primary
variable in determining the light-curve properties of a particular SN,
with younger systems located at higher-redshift when compared to those
seen locally. It is important to note that these potential trends will
not affect the validity of the cosmological result. Larger samples of
SNe are clearly required to probe these trends further.
  
We find weak trends between light-curve shape and the spatial location
of the SNe in the host galaxy, as claimed occasionally in lower
redshift samples, in the sense that fainter SNe are located away from
galaxy centres. However, there are two caveats associated with this
result: firstly, there is a selection effect of faint SNe being missed
near to galaxy cores at low-redshift \citep[this effect is not
believed to operate to the same extent at high-redshift, see
e.g.][]{1996ApJ...473..356P}, and secondly, we use projected distances
rather than de-projected distances in the analysis.

We interpret the above results according to the hypothesis that dust
extinction in the host galaxy is the significant cause of the scatter
in the Hubble diagram. The implications of this conclusion are
significant. Most importantly, \textit{the Hubble diagram confined to
  SNe\,Ia occurring in early-type host galaxies with presumed minimal
  internal extinction presents a very tight relationship and provides
  a $\simeq5\sigma$ confirmation of a non-zero cosmological constant,
  assuming a flat Universe.} When no assumption about the flatness of
the Universe are made, we find that SNe in E/S0 galaxies imply $\ol>0$
at nearly 98 per cent probability.

The differential trends in extinction by host galaxy morphology as
revealed by the SN colour excess and mean offset in the type-dependent
Hubble diagram with respect to that defined by the early-type hosts
implies only modest extinctions in late-type spirals of
$A_B=0.14\pm0.09$\,mag, though with a large galaxy-to-galaxy scatter.
We discuss briefly the origin of this small extinction in the light of
the increasing fraction of star-forming and presumed dusty galaxies at
high redshift and suggest a number of possible explanations.

Our study is a good illustration of how ancillary data on host
galaxies can be used to examine the nature and reliability of SNe as
probes of the cosmological expansion history. Contrary to suggestions
made by some previous studies \citep[e.g.][]{2002MNRAS.332..352R}, we
show that internal dust extinction cannot be a primary contaminant and
that supernovae of Type Ia, particularly those occurring in spheroidal
galaxies which can be readily screened with spectroscopic, colour and
morphological data, represent a very powerful cosmological probe. This
conclusion is of particular significance for future projects such as
the \textit{SuperNova Acceleration Probe (SNAP)} experiment (e.g.
Aldering et al. 2002), a space-based telescope designed to determine
the cosmic expansion history to high precision via studies of many
thousands of SNe\,Ia. This instrument will provide \hst-quality
imaging and nine-band optical and near-infrared colour photometry for
the host galaxies of nearly every SN it discovers, allowing this
host-galaxy study to be extended using better statistics and a
superior classification of the host-galaxy types, and a particularly
robust cosmological solution from those events located in spheroidal
systems.

\section*{Acknowledgements}

The study is based on observations made with the NASA/ESA
\textit{Hubble Space Telescope}, obtained at the Space Telescope
Science Institute, which is operated by the Association of
Universities for Research in Astronomy, Inc., under NASA contract NAS
5-26555. These observations are associated with proposal numbers
8313/9131. Some of the data presented herein were obtained at the W.M.
Keck Observatory, which is operated as a scientific partnership among
the California Institute of Technology, the University of California
and the National Aeronautics and Space Administration. The Observatory
was made possible by the generous financial support of the W.M. Keck
Foundation. The authors wish to recognise and acknowledge the very
significant cultural role and reverence that the summit of Mauna Kea
has always had within the indigenous Hawaiian community.  We are most
fortunate to have the opportunity to conduct observations from this
mountain. MS would like to acknowledge generous financial support from
a PPARC fellowship. We thank David Bacon, Richard Massey and Pranjal
Trivedi for taking certain spectroscopic and photometric observations
at the Keck-II and Palomar 200-inch telescopes. This work is supported
in part by the Physics Division, E.  O. Lawrence Berkeley National
Laboratory of the U.S. Department of Energy under contract
DE-AC03-76SF000098, and by the Center for Particle Astrophysics, an
NSF Science and Technology Center operated by the University of
California, Berkeley, under Cooperative Agreement No.  AST-91-20005.

\bibliographystyle{apj}

\begin{thebibliography}{68}
\expandafter\ifx\csname natexlab\endcsname\relax\def\natexlab#1{#1}\fi

\bibitem[{{Aguirre}(1999)}]{1999ApJ...525..583A}
{Aguirre}, A. 1999, \apj, 525, 583

\bibitem[{{Aguirre} \& {Haiman}(2000)}]{2000ApJ...532...28A}
{Aguirre}, A. \& {Haiman}, Z. 2000, \apj, 532, 28

\bibitem[{{Aldering} {et~al.}(2002){Aldering}, {et al.}}]{alderingSPIE}{Aldering}, G., et al., 2002, in Dressler, A., ed, \procspie, Vol. 4835, Future Research Direction and Visions for Astronomy, SPIE Press 

\bibitem[{{Bahcall} {et~al.}(1999){Bahcall}, {Ostriker}, {Perlmutter}, \&
  {Steinhardt}}]{1999Sci...284.1481B}
{Bahcall}, N.~A., {Ostriker}, J.~P., {Perlmutter}, S., \& {Steinhardt}, P.~J.
  1999, Science, 284, 1481

\bibitem[{{Balbi} {et~al.}(2000){Balbi}, {Ade}, {Bock}, {Borrill}, {Boscaleri},
  {De Bernardis}, {Ferreira}, {Hanany}, {Hristov}, {Jaffe}, {Lee}, {Oh},
  {Pascale}, {Rabii}, {Richards}, {Smoot}, {Stompor}, {Winant}, \&
  {Wu}}]{2000ApJ...545L...1B}
{Balbi}, A., et al. 2000,
  \apjl, 545, L1

\bibitem[{{Bertin} \& {Arnouts}(1996)}]{1996A&AS..117..393B}
{Bertin}, E. \& {Arnouts}, S. 1996, \aaps, 117, 393

\bibitem[{{Branch} {et~al.}(1996){Branch}, {Romanishin}, \&
  {Baron}}]{1996ApJ...465...73B}
{Branch}, D., {Romanishin}, W., \& {Baron}, E. 1996, \apj, 465, 73

\bibitem[{{Brinchmann} {et~al.}(1998){Brinchmann}, {Abraham}, {Schade},
  {Tresse}, {Ellis}, {Lilly}, {Le Fevre}, {Glazebrook}, {Hammer}, {Colless},
  {Crampton}, \& {Broadhurst}}]{1998ApJ...499..112B}
{Brinchmann}, J., et al. 1998, \apj, 499, 112

\bibitem[{{Brinchmann} \& {Ellis}(2000)}]{2000ApJ...536L..77B}
{Brinchmann}, J. \& {Ellis}, R.~S. 2000, \apjl, 536, L77

\bibitem[{{Calzetti} {et~al.}(2000){Calzetti}, {Armus}, {Bohlin}, {Kinney},
  {Koornneef}, \& {Storchi-Bergmann}}]{2000ApJ...533..682C}
{Calzetti}, D., {Armus}, L., {Bohlin}, R.~C., {Kinney}, A.~L., {Koornneef}, J.,
  \& {Storchi-Bergmann}, T. 2000, \apj, 533, 682

\bibitem[{{Calzetti} {et~al.}(1994){Calzetti}, {Kinney}, \&
  {Storchi-Bergmann}}]{1994ApJ...429..582C}
{Calzetti}, D., {Kinney}, A.~L., \& {Storchi-Bergmann}, T. 1994, \apj, 429, 582

\bibitem[{{Cardelli} {et~al.}(1989){Cardelli}, {Clayton}, \&
  {Mathis}}]{1989ApJ...345..245C}
{Cardelli}, J.~A., {Clayton}, G.~C., \& {Mathis}, J.~S. 1989, \apj, 345, 245

\bibitem[{{de Bernardis} {et~al.}(2002){de Bernardis}, {Ade}, {Bock}, {Bond},
  {Borrill}, {Boscaleri}, {Coble}, {Contaldi}, {Crill}, {De Troia}, {Farese},
  {Ganga}, {Giacometti}, {Hivon}, {Hristov}, {Iacoangeli}, {Jaffe}, {Jones},
  {Lange}, {Martinis}, {Masi}, {Mason}, {Mauskopf}, {Melchiorri}, {Montroy},
  {Netterfield}, {Pascale}, {Piacentini}, {Pogosyan}, {Polenta}, {Pongetti},
  {Prunet}, {Romeo}, {Ruhl}, \& {Scaramuzzi}}]{2002ApJ...564..559D}
{de Bernardis}, P., et al. 2002, \apj, 564, 559

\bibitem[{{de Bernardis} {et~al.}(2000){de Bernardis}, {Ade}, {Bock}, {Bond},
  {Borrill}, {Boscaleri}, {Coble}, {Crill}, {De Gasperis}, {Farese},
  {Ferreira}, {Ganga}, {Giacometti}, {Hivon}, {Hristov}, {Iacoangeli}, {Jaffe},
  {Lange}, {Martinis}, {Masi}, {Mason}, {Mauskopf}, {Melchiorri}, {Miglio},
  {Montroy}, {Netterfield}, {Pascale}, {Piacentini}, {Pogosyan}, {Prunet},
  {Rao}, {Romeo}, {Ruhl}, {Scaramuzzi}, {Sforna}, \&
  {Vittorio}}]{2000Natur.404..955D}
{de Bernardis}, P., et al. 2000, \nat,
  404, 955

\bibitem[{{Dom{\' i}nguez} {et~al.}(2001){Dom{\' i}nguez}, {H{\" o}flich}, \&
  {Straniero}}]{2001ApJ...557..279D}
{Dom{\' i}nguez}, I., {H{\" o}flich}, P., \& {Straniero}, O. 2001, \apj, 557,
  279

\bibitem[{{Efstathiou} {et~al.}(2002){Efstathiou}, {Moody}, {Peacock},
  {Percival}, {Baugh}, {Bland-Hawthorn}, {Bridges}, {Cannon}, {Cole},
  {Colless}, {Collins}, {Couch}, {Dalton}, {de Propris}, {Driver}, {Ellis},
  {Frenk}, {Glazebrook}, {Jackson}, {Lahav}, {Lewis}, {Lumsden}, {Maddox},
  {Norberg}, {Peterson}, {Sutherland}, \& {Taylor}}]{2002MNRAS.330L..29E}
{Efstathiou}, G., et al. 2002, \mnras, 330, L29

\bibitem[{{Fanelli} {et~al.}(1988){Fanelli}, {O'Connell}, \&
  {Thuan}}]{1988ApJ...334..665F}
{Fanelli}, M.~N., {O'Connell}, R.~W., \& {Thuan}, T.~X. 1988, \apj, 334, 665

\bibitem[{{Filippenko} \& {Sargent}(1989)}]{1989ApJ...345L..43F}
{Filippenko}, A.~V. \& {Sargent}, W.~L.~W. 1989, \apjl, 345, L43

\bibitem[{{Fruchter} \& {Hook}(1997)}]{1997SPIE.3164..120F}
{Fruchter}, A. \& {Hook}, R.~N. 1997, in Proc. SPIE Vol. 3164, p. 120-125,
  Applications of Digital Image Processing XX, Andrew G. Tescher; Ed., Vol.
  3164, 120--125

\bibitem[{{Gal-Yam} {et~al.}(2002){Gal-Yam}, {Maoz}, \&
  {Sharon}}]{2002MNRAS.332...37G}
{Gal-Yam}, A., {Maoz}, D., \& {Sharon}, K. 2002, \mnras, 332, 37

\bibitem[{{Goldhaber} {et~al.}(2001){Goldhaber}, {Groom}, {Kim}, {Aldering},
  {Astier}, {Conley}, {Deustua}, {Ellis}, {Fabbro}, {Fruchter}, {Goobar},
  {Hook}, {Irwin}, {Kim}, {Knop}, {Lidman}, {McMahon}, {Nugent}, {Pain},
  {Panagia}, {Pennypacker}, {Perlmutter}, {Ruiz-Lapuente}, {Schaefer},
  {Walton}, \& {York}}]{2001ApJ...558..359G}
{Goldhaber}, G., et al. 2001, \apj, 558, 359

\bibitem[{{Goobar} \& {Perlmutter}(1995)}]{1995ApJ...450...14G}
{Goobar}, A. \& {Perlmutter}, S. 1995, \apj, 450, 14

\bibitem[{{H{\" o}flich} {et~al.}(2000){H{\" o}flich}, {Nomoto}, {Umeda}, \&
  {Wheeler}}]{2000ApJ...528..590H}
{H{\" o}flich}, P., {Nomoto}, K., {Umeda}, H., \& {Wheeler}, J.~C. 2000, \apj,
  528, 590

\bibitem[{{Hamuy} {et~al.}(1995){Hamuy}, {Phillips}, {Maza}, {Suntzeff},
  {Schommer}, \& {Aviles}}]{1995AJ....109....1H}
{Hamuy}, M., {Phillips}, M.~M., {Maza}, J., {Suntzeff}, N.~B., {Schommer},
  R.~A., \& {Aviles}, R. 1995, \aj, 109, 1

\bibitem[{{Hamuy} {et~al.}(1996{\natexlab{a}}){Hamuy}, {Phillips}, {Suntzeff},
  {Schommer}, {Maza}, {Antezan}, {Wischnjewsky}, {Valladares}, {Muena},
  {Gonzales}, {Aviles}, {Wells}, {Smith}, {Navarrete}, {Covarrubias},
  {Williger}, {Walker}, {Layden}, {Elias}, {Baldwin}, {Hernandez}, {Tirado},
  {Ugarte}, {Elston}, {Saavedra}, {Barrientos}, {Costa}, {Lira}, {Ruiz},
  {Anguita}, {Gomez}, {Ortiz}, {della Valle}, {Danziger}, {Storm}, {Kim},
  {Bailyn}, {Rubenstein}, {Tucker}, {Cersosimo}, {Mendez}, {Siciliano},
  {Sherry}, {Chaboyer}, {Koopmann}, {Geisler}, {Sarajedini}, {Dey}, {Tyson},
  {Rich}, {Gal}, {Lamontagne}, {Caldwell}, {Guhathakurta}, {Phillips},
  {Szkody}, {Prosser}, {Ho}, {McMahan}, {Baggley}, {Cheng}, {Havlen},
  {Wakamatsu}, {Janes}, {Malkan}, {Baganoff}, {Seitzer}, {Shara}, {Sturch},
  {Hesser}, {Hartig}, {Hughes}, {Welch}, {Williams}, {Ferguson}, {Francis},
  {French}, {Bolte}, {Roth}, {Odewahn}, {Howell}, \&
  {Krzeminski}}]{1996AJ....112.2408H}
{Hamuy}, M., et al. 1996{\natexlab{a}}, \aj, 112, 2408

\bibitem[{{Hamuy} {et~al.}(1996{\natexlab{b}}){Hamuy}, {Phillips}, {Suntzeff},
  {Schommer}, {Maza}, \& {Aviles}}]{1996AJ....112.2391H}
{Hamuy}, M., {Phillips}, M.~M., {Suntzeff}, N.~B., {Schommer}, R.~A., {Maza},
  J., \& {Aviles}, R. 1996{\natexlab{b}}, \aj, 112, 2391

\bibitem[{{Hamuy} {et~al.}(1996{\natexlab{c}}){Hamuy}, {Phillips}, {Suntzeff},
  {Schommer}, {Maza}, \& {Aviles}}]{1996AJ....112.2398H}
---. 1996{\natexlab{c}}, \aj, 112, 2398

\bibitem[{{Hamuy} {et~al.}(2000){Hamuy}, {Trager}, {Pinto}, {Phillips},
  {Schommer}, {Ivanov}, \& {Suntzeff}}]{2000AJ....120.1479H}
{Hamuy}, M., {Trager}, S.~C., {Pinto}, P.~A., {Phillips}, M.~M., {Schommer},
  R.~A., {Ivanov}, V., \& {Suntzeff}, N.~B. 2000, \aj, 120, 1479

\bibitem[{{Hatano} {et~al.}(1998){Hatano}, {Branch}, \&
  {Deaton}}]{1998ApJ...502..177H}
{Hatano}, K., {Branch}, D., \& {Deaton}, J. 1998, \apj, 502, 177

\bibitem[{{Howell}(2001)}]{2001ApJ...554L.193H}
{Howell}, D.~A. 2001, \apjl, 554, L193

\bibitem[{{Howell} {et~al.}(2000){Howell}, {Wang}, \&
  {Wheeler}}]{2000ApJ...530..166H}
{Howell}, D.~A., {Wang}, L., \& {Wheeler}, J.~C. 2000, \apj, 530, 166

\bibitem[{{Ivanov} {et~al.}(2000){Ivanov}, {Hamuy}, \&
  {Pinto}}]{2000ApJ...542..588I}
{Ivanov}, V.~D., {Hamuy}, M., \& {Pinto}, P.~A. 2000, \apj, 542, 588

\bibitem[{{Jaffe} {et~al.}(2001){Jaffe}, {Ade}, {Balbi}, {Bock}, {Bond},
  {Borrill}, {Boscaleri}, {Coble}, {Crill}, {de Bernardis}, {Farese},
  {Ferreira}, {Ganga}, {Giacometti}, {Hanany}, {Hivon}, {Hristov},
  {Iacoangeli}, {Lange}, {Lee}, {Martinis}, {Masi}, {Mauskopf}, {Melchiorri},
  {Montroy}, {Netterfield}, {Oh}, {Pascale}, {Piacentini}, {Pogosyan},
  {Prunet}, {Rabii}, {Rao}, {Richards}, {Romeo}, {Ruhl}, {Scaramuzzi},
  {Sforna}, {Smoot}, {Stompor}, {Winant}, \& {Wu}}]{2001PhRvL..86.3475J}
{Jaffe}, A.~H., et al. 2001, Physical Review Letters,
  86, 3475

\bibitem[{{Jorgensen}(1994)}]{1994PASP..106..967J}
{Jorgensen}, I. 1994, \pasp, 106, 967

\bibitem[{{Kim} {et~al.}(1996){Kim}, {Goobar}, \&
  {Perlmutter}}]{1996PASP..108..190K}
{Kim}, A., {Goobar}, A., \& {Perlmutter}, S. 1996, \pasp, 108, 190

\bibitem[{{Krisciunas}(1990)}]{1990PASP..102.1052K}
{Krisciunas}, K. 1990, \pasp, 102, 1052

\bibitem[{{Krisciunas} {et~al.}(1987){Krisciunas}, {Sinton}, {Tholen},
  {Tokunaga}, {Golisch}, {Griep}, {Kaminski}, {Impey}, \&
  {Christian}}]{1987PASP...99..887K}
{Krisciunas}, K., et al. 1987, \pasp, 99,
  887

\bibitem[{{Kron}(1980)}]{1980ApJS...43..305K}
{Kron}, R.~G. 1980, \apjs, 43, 305

\bibitem[{{Landolt}(1992)}]{1992AJ....104..340L}
{Landolt}, A.~U. 1992, \aj, 104, 340

\bibitem[{{Lilly} {et~al.}(1996){Lilly}, {Le Fevre}, {Hammer}, \&
  {Crampton}}]{1996ApJ...460L...1L}
{Lilly}, S.~J., {Le Fevre}, O., {Hammer}, F., \& {Crampton}, D. 1996, \apjl,
  460, L1

\bibitem[{{Mas-Hesse} \& {Kunth}(1999)}]{1999A&A...349..765M}
{Mas-Hesse}, J.~M. \& {Kunth}, D. 1999, \aap, 349, 765

\bibitem[{{Massey} \& {Gronwall}(1990)}]{1990ApJ...358..344M}
{Massey}, P. \& {Gronwall}, C. 1990, \apj, 358, 344

\bibitem[{{Massey} {et~al.}(1988){Massey}, {Strobel}, {Barnes}, \&
  {Anderson}}]{1988ApJ...328..315M}
{Massey}, P., {Strobel}, K., {Barnes}, J.~V., \& {Anderson}, E. 1988, \apj,
  328, 315

\bibitem[{{Nugent} {et~al.}(2002){Nugent}, {Kim}, \&
  {Perlmutter}}]{2002PASP..114..803N}
{Nugent}, P., {Kim}, A., \& {Perlmutter}, S. 2002, \pasp, 114, 803

\bibitem[{{Osterbrock}(1989)}]{1989agna.book.....O}
{Osterbrock}, D.~E. 1989, Astrophysics of gaseous nebulae and active galactic
  nuclei (Mill Valley, CA, University Science Books, 1989, 422 p.)

\bibitem[{{Pain} {et~al.}(2002){Pain}, {Fabbro}, {Sullivan}, {Ellis},
  {Aldering}, {Astier}, {Duestua}, {Fruchter}, {Goldhaber}, {Goobar}, {Groom},
  {Hardin}, {Hook}, {Howell}, {Irwin}, {Kim}, {Kim}, {Knop}, {Lee}, {Lidman},
  {McMahon}, {Nugent}, {Panagia}, {Pennypacker}, {Perlmutter}, {Ruiz-Lapuente},
  {Schahmaneche}, {Schaefer}, \& {Walton.}}]{2002astro.ph..5476P}
{Pain}, R., et al. 2002, in press, \apj, astroph/0205476

\bibitem[{{Pain} {et~al.}(1996){Pain}, {Hook}, {Deustua}, {Gabi}, {Goldhaber},
  {Groom}, {Kim}, {Kim}, {Lee}, {Pennypacker}, {Perlmutter}, {Small}, {Goobar},
  {Ellis}, {McMahon}, {Glazebrook}, {Boyle}, {Bunclark}, {Carter}, \&
  {Irwin}}]{1996ApJ...473..356P}
{Pain}, R., et al. 1996, \apj, 473, 356

\bibitem[{{Peacock} {et~al.}(2001){Peacock}, {Cole}, {Norberg}, {Baugh},
  {Bland-Hawthorn}, {Bridges}, {Cannon}, {Colless}, {Collins}, {Couch},
  {Dalton}, {Deeley}, {De Propris}, {Driver}, {Efstathiou}, {Ellis}, {Frenk},
  {Glazebrook}, {Jackson}, {Lahav}, {Lewis}, {Lumsden}, {Maddox}, {Percival},
  {Peterson}, {Price}, {Sutherland}, \& {Taylor}}]{2001Natur.410..169P}
{Peacock}, J.~A., et al. 2001, \nat, 410, 169

\bibitem[{{Pei}(1992)}]{1992ApJ...395..130P}
{Pei}, Y.~C. 1992, \apj, 395, 130

\bibitem[{{Perlmutter}(1997)}]{1997perlbook..749}
{Perlmutter}, S. 1997, in Thermonuclear Supernovae, ed. P. Ruiz-Lapuente, R.
  Canal, \& J. Isern (NATO ASI Ser. C, 486) (Dordrecht: Kluwer), 749

\bibitem[{{Perlmutter} {et~al.}(1998){Perlmutter}, {Aldering}, {della Valle},
  {Deustua}, {Ellis}, {Fabbro}, {Fruchter}, {Goldhaber}, {Groom}, {Hook},
  {Kim}, {Kim}, {Knop}, {Lidman}, {McMahon}, {Nugent}, {Pain}, {Panagia},
  {Pennypacker}, {Ruiz-Lapuente}, {Schaefer}, \&
  {Walton}}]{1998Natur.391...51P}
{Perlmutter}, S., et al. 1998, \nat, 391, 51

\bibitem[{{Perlmutter} {et~al.}(1999){Perlmutter}, {Aldering}, {Goldhaber},
  {Knop}, {Nugent}, {Castro}, {Deustua}, {Fabbro}, {Goobar}, {Groom}, {Hook},
  {Kim}, {Kim}, {Lee}, {Nunes}, {Pain}, {Pennypacker}, {Quimby}, {Lidman},
  {Ellis}, {Irwin}, {McMahon}, {Ruiz-Lapuente}, {Walton}, {Schaefer}, {Boyle},
  {Filippenko}, {Matheson}, {Fruchter}, {Panagia}, {Newberg}, {Couch}, \& {The
  Supernova Cosmology Project}}]{1999ApJ...517..565P}
{Perlmutter}, S., et al. 1999, \apj, 517, 565 (P99)

\bibitem[{{Perlmutter} {et~al.}(1997){Perlmutter}, {Gabi}, {Goldhaber},
  {Goobar}, {Groom}, {Hook}, {Kim}, {Kim}, {Lee}, {Pain}, {Pennypacker},
  {Small}, {Ellis}, {McMahon}, {Boyle}, {Bunclark}, {Carter}, {Irwin},
  {Glazebrook}, {Newberg}, {Filippenko}, {Matheson}, {Dopita}, {Couch}, \& {The
  Supernova Cosmology Project}}]{1997ApJ...483..565P}
{Perlmutter}, S., et al. 1997,
  \apj, 483, 565 (P97)

\bibitem[{{Phillips}(1993)}]{1993ApJ...413L.105P}
{Phillips}, M.~M. 1993, \apjl, 413, L105

\bibitem[{{Phillips} {et~al.}(1999){Phillips}, {Lira}, {Suntzeff}, {Schommer},
  {Hamuy}, \& {Maza}}]{1999AJ....118.1766P}
{Phillips}, M.~M., {Lira}, P., {Suntzeff}, N.~B., {Schommer}, R.~A., {Hamuy},
  M., \& {Maza}, J.~. 1999, \aj, 118, 1766

\bibitem[{{Poggianti}(1997)}]{1997A&AS..122..399P}
{Poggianti}, B.~M. 1997, \aaps, 122, 399

\bibitem[{{Riess} {et~al.}(1998){Riess}, {Filippenko}, {Challis},
  {Clocchiatti}, {Diercks}, {Garnavich}, {Gilliland}, {Hogan}, {Jha},
  {Kirshner}, {Leibundgut}, {Phillips}, {Reiss}, {Schmidt}, {Schommer},
  {Smith}, {Spyromilio}, {Stubbs}, {Suntzeff}, \&
  {Tonry}}]{1998AJ....116.1009R}
{Riess}, A.~G., et al. 1998, \aj, 116, 1009

\bibitem[{{Riess} {et~al.}(1999){Riess}, {Kirshner}, {Schmidt}, {Jha},
  {Challis}, {Garnavich}, {Esin}, {Carpenter}, {Grashius}, {Schild}, {Berlind},
  {Huchra}, {Prosser}, {Falco}, {Benson}, {Brice{\~ n}o}, {Brown}, {Caldwell},
  {dell'Antonio}, {Filippenko}, {Goodman}, {Grogin}, {Groner}, {Hughes},
  {Green}, {Jansen}, {Kleyna}, {Luu}, {Macri}, {McLeod}, {McLeod}, {McNamara},
  {McLean}, {Milone}, {Mohr}, {Moraru}, {Peng}, {Peters}, {Prestwich},
  {Stanek}, {Szentgyorgyi}, \& {Zhao}}]{1999AJ....117..707R}
{Riess}, A.~G., et al. 1999, \aj, 117, 707

\bibitem[{{Riess} {et~al.}(1995){Riess}, {Press}, \&
  {Kirshner}}]{1995ApJ...438L..17R}
{Riess}, A.~G., {Press}, W.~H., \& {Kirshner}, R.~P. 1995, \apjl, 438, L17

\bibitem[{{Riess} {et~al.}(1996){Riess}, {Press}, \&
  {Kirshner}}]{1996ApJ...473...88R}
---. 1996, \apj, 473, 88

\bibitem[{{Rowan-Robinson}(2002)}]{2002MNRAS.332..352R}
{Rowan-Robinson}, M. 2002, \mnras, 332, 352

\bibitem[{{Ruiz-Lapuente} {et~al.}(1997){Ruiz-Lapuente}, {Canal}, \& {Burkert}}]{ruizlapuente97}
{Ruiz-Lapuente}, P., {Canal}, R., \& {Burkert}, A. 1997, 
in Thermonuclear Supernovae, ed. P. Ruiz-Lapuente, R.
  Canal, \& J. Isern (NATO ASI Ser. C, 486) (Dordrecht: Kluwer), 205

\bibitem[{{Schlegel} {et~al.}(1998){Schlegel}, {Finkbeiner}, \&
  {Davis}}]{1998ApJ...500..525S}
{Schlegel}, D.~J., {Finkbeiner}, D.~P., \& {Davis}, M. 1998, \apj, 500, 525

\bibitem[{{Schmidt} {et~al.}(1998){Schmidt}, {Suntzeff}, {Phillips},
  {Schommer}, {Clocchiatti}, {Kirshner}, {Garnavich}, {Challis}, {Leibundgut},
  {Spyromilio}, {Riess}, {Filippenko}, {Hamuy}, {Smith}, {Hogan}, {Stubbs},
  {Diercks}, {Reiss}, {Gilliland}, {Tonry}, {Maza}, {Dressler}, {Walsh}, \&
  {Ciardullo}}]{1998ApJ...507...46S}
{Schmidt}, B.~P., et al. 1998, \apj, 507, 46

\bibitem[{{Shaw}(1979)}]{1979A&A....76..188S}
{Shaw}, R.~L. 1979, \aap, 76, 188

\bibitem[{{Sheinis} {et~al.}(2002){Sheinis}, {Bolte}, {Epps}, {Kibrick},
  {Miller}, {Radovan}, {Bigelow}, \& {Sutin}}]{2002PASP..114..851S}
{Sheinis}, A.~I., {Bolte}, M., {Epps}, H.~W., {Kibrick}, R.~I., {Miller},
  J.~S., {Radovan}, M.~V., {Bigelow}, B.~C., \& {Sutin}, B.~M. 2002, \pasp,
  114, 851

\bibitem[{{Thuan} \& {Gunn}(1976)}]{1976PASP...88..543T}
{Thuan}, T.~X. \& {Gunn}, J.~E. 1976, \pasp, 88, 543

\bibitem[{{Totani} \& {Kobayashi}(1999)}]{1999ApJ...526L..65T}
{Totani}, T. \& {Kobayashi}, C. 1999, \apjl, 526, L65

\bibitem[{{Umeda} {et~al.}(1999){Umeda}, {Nomoto}, {Kobayashi}, {Hachisu}, \&
  {Kato}}]{1999ApJ...522L..43U}
{Umeda}, H., {Nomoto}, K., {Kobayashi}, C., {Hachisu}, I., \& {Kato}, M. 1999,
  \apjl, 522, L43

\bibitem[{{Wang} {et~al.}(1997){Wang}, {Hoeflich}, \&
  {Wheeler}}]{1997ApJ...483L..29W}
{Wang}, L., {Hoeflich}, P., \& {Wheeler}, J.~C. 1997, \apjl, 483, L29

\end{thebibliography}

\end{document}